\documentclass[11pt,reqno]{amsart}
\usepackage{amsmath,amsgen,amstext,amsbsy,amsopn,amsthm,amssymb}

\newtheorem{thm}{THEOREM}

\newtheorem{lem}{LEMMA}

\theoremstyle{definition}
\newtheorem{definition}{DEFINITION}

\newcommand{\R}{{\mathbb R}}

\renewcommand{\c}{{\cdot}}

\newcommand{\D}{{D}_{\mathrm{f}}}
\newcommand{\Di}{{D}^{-1}_{\mathrm{f}}}
\newcommand{\Dh}{{D}^{-1/2}_{\mathrm{f}}}
\newcommand{\eat}{E_{\mathrm{at}}}
\renewcommand{\P}{{P}_{\mathrm{f}}}
\newcommand{\bP}{{\bar P}}
\newcommand{\bH}{{\bar H}}
\newcommand{\Hf}{{H}_{\mathrm{f}}}
\newcommand{\F}{{\mathcal F}}
\newcommand{\Li}{{L^{-1}}}
\newcommand{\Lh}{{L^{-1/2}}}

\newcommand{\dv}{{R}}

\renewcommand{\H}{{\mathcal H}}

\newcommand{\eps}{\varepsilon}

\newcommand{\vp}{\varphi}

\newcommand{\Ow}{{\mathcal O}}
\renewcommand{\o}{{\Omega}}
\numberwithin{equation}{section}

\title[Binding energy without cutoffs]{Binding energy for hydrogen-like atoms in the Nelson model without 
cutoffs}

\thanks{ C.H.
    acknowledges support through the European Union's IHP
    network Analysis \& Quantum HPRN-CT-2002-00277. M.H. 
    is supported by JSPS, Grant-in-Aid for Scientific Research
    (C) 13640215. C.H. and H.S. are thankful to I. Catto for useful comments.}

\author[C. Hainzl]{Christian Hainzl}
\address{CEREMADE, Universit\'e
  Paris-Dauphine, Mar\'echal de Tassigny, F-75775 Paris
  Cedex 16, \& Laboratoire de Math\'ematiques
Paris-Sud-Bat 425, F-91405 Orsay Cedex, France.} \email{hainzl@ceremade.dauphine.fr}

\author[M. Hirokawa]{Masao Hirokawa}
 \address{Department of Mathematics, Okayama University, 7000-8530, Okayama, Japan}
  \email{hirokawa@math.okayama-u.ac.jp}

\author[H. Spohn]{Herbert Spohn}
 \address{Technische Universit\"at M\"unchen, Zentrum Mathematik, 85747 Garching, Germany}
  \email{spohn@ma.tum.de}

\begin{document}

\begin{abstract}
In the Nelson model particles interact through a scalar massless field.
For hydrogen-like atoms there is a nucleus of infinite mass and charge
$Ze$, $Z > 0$, fixed at the origin and an electron of mass $m$ and charge $e$.
This system forms a bound state with binding energy $E_{\rm bin} = me^4Z^2/2$ 
to leading order in $e$. We investigate the radiative corrections 
to the binding energy and prove upper and lower bounds
which imply that $ E_{\rm bin} = me^4Z^2/2 + c_0 e^6 + \Ow(e^7 \ln e)$
with explicit coefficient $c_0$ and independent of the
ultraviolet cutoff. $c_0$ can be computed by 
perturbation theory, which however is only formal since
for the Nelson Hamiltonian the smallest eigenvalue sits exactly at the bottom of the 
continuous spectrum.
\end{abstract}
\maketitle

\section{Introduction}

As a very famous result in the early days of constructive quantum
field theory, E. Nelson proved that for charges coupled to a
scalar massless Bose field the ultraviolet cutoff can be removed at the
expense of an infinite energy renormalization \cite{N}. In our contribution
we study Nelson's model for the case of a
hydrogen-like atom. It consists of a nucleus of infinite mass,
nailed down at the origin, carrying a charge $Ze$, $Z>0$, and a
quantum particle, called electron for simplicity, of mass $m$ carrying
charge $e$. Note that in the Nelson model charges of equal
sign attract each other. Without restriction we may set $e \geq
0$. We also use natural units, for which $m=1/2$, $\hbar=1$,
$c=1$. Thus $e$ remains as the only parameter in the model. Our
goal is to obtain precise estimates on the binding energy and thus
to prove that Nelson's renormalized Hamiltonian is in agreement
with the experimental fact of small radiative corrections.

On the physical level the binding energy is computed formally through
perturbation theory. Since in the Nelson Hamiltonian the 
ground state is not separated by a gap
from the continuous spectrum, there is no hope to justify such a procedure
mathematically \`{a} la Kato \cite{K}. In fact, as to be shown, the
binding energy is not analytic in $e$ near $e=0$. To have a
substitute for the formal perturbation theory, we will develop a
method which yields upper and lower bounds on the binding energy.
In principle, our scheme can be pushed to arbitrary order. In the
present contribution we include the first radiative correction of
order $e^6$ with an error $\mathcal{O}(e^7 \log e)$. In
\cite{H1,CH} a similar scheme has been established. Here we
advance in two central issues. Firstly for the binding energy the
iteration scheme has to incorporate an external potential.
Secondly, we have to make sure that the bounds are uniform in the
ultraviolet cutoff. As an extra bonus, the principles underlying the
theory in \cite{CH} are stated more clearly and we believe that in
the present form the iteration scheme can be applied directly to
other models of a similar structure.

The Hilbert space for the electron is $L^2(\R^3)=
\mathcal{H}_{\mathrm{p}}$ and the one for the scalar Bose field
the symmetric Fock space $\mathcal{F}$ over $L^2(\R^3)$ as
one-particle space. The coupled system has $\mathcal{H}=
\mathcal{H}_{\mathrm{p}} \otimes \mathcal{F}$ as state space. Its
scalar product is denoted by $(\cdot,\cdot)_{\mathcal{H}}$. We
omit the index if obvious from the context. On $L^2(\R^3)$ the
canonical pair is the multiplication operator $x$ and $-i p=
\nabla_x$. The Bose field on $\mathcal{F}$ is given through the
creation and annihilation operators, $a^*(f)$, $a(g)$, which are
densely defined for test functions $f,g \in L^2(\R^3)$. The field
satisfies the canonical commutation relations
\begin{equation}
[a(f),a^*(g)]= (f,g)_{L^2}\,,\quad [a(f),a(g)]=0\,,\quad
[a^*(f),a^*(g)]=0\,.
\end{equation}
With this notation the Hamiltonian for the particle reads
\begin{equation}
H_{\mathrm{at}}= p^2- \frac{Ze^2}{4\pi|x|}
\end{equation}
and the one for the field is given by
\begin{equation}
H_{\mathrm{f}}= \int dk \omega(k)a^*(k)a(k)\,,\quad \omega(k)=|k|
\,,
\end{equation}
i.e., $H_{\mathrm{f}}$ is the second quantization of $\omega$
considered as multiplication operator on $L^2(\R^3,dk)$.

The coupling is mediated through $a(\vp)$, $a^*(\vp)$ with a special
choice of $\varphi$. Notationally it is convenient to have a
distinguished symbol. We set
\begin{equation}\label{4}
\begin{split}
& A= \int dk \chi(k) \frac{1}{\sqrt{2\omega}}
\frac{k}{\omega+k^2} a(k)\,, \\
& A= a(\vp)\quad\mathrm{with}\quad
\vp=\chi(k)\frac{1}{\sqrt{2\omega}}\frac{k}{\omega+k^2} \,,
\end{split}
\end{equation}
where $\chi(k)=(2\pi)^{-3/2}$ for $|k|< \Lambda$ and
$\chi(k)=0$ for $|k|> \Lambda$. For $\Lambda = \infty$ the test
function appearing in \eqref{4} fails to be in $L^2$ because of
logarithmic divergence at $k=\infty$. Thus we have to keep
$\Lambda < \infty$ in intermediate steps and make sure that all
estimates are uniform in $\Lambda$. $A$ is a 3-vector. In
expressions as $A^*A$, $pA$ we really should write $A^*\cdot A$,
$p\cdot A$. Our convention is that in strings of 3-vectors the
scalar product is understood in pairs, e.g., $AAA^*p$ means
$(A\cdot A)(A^*\cdot p)$.

The renormalized Nelson Hamiltonian $H_{\mathrm{ren}}$ in case of
hydrogen-like atoms is explained in \cite{HHS}. We use
$H_{\mathrm{ren}}$ as our starting point with the small
modification that $H_{\mathrm{ren}}$ is unitarily transformed to
\begin{equation}
H= e^{ix \P} H_{\mathrm{ren}}e^{-ix \P}\,,
\end{equation}
where $\P$ denotes the momentum of the Bose field,
\begin{equation}
\P= \int dk \, ka^*(k)a(k)\,.
\end{equation}
Physically, $p$ acquires then the meaning of the total momentum,
i.e., the momentum of particle + field, rather than the momentum of
the particle by itself. The starting Hamiltonian is thus, for $Z
\geq 0$,
\begin{multline}\label{6}
H_\Lambda = p^2 -\frac{Ze^2}{4\pi|x|} + \Hf +\P^2 - 2p  \P -
2e\big(A^*(p-\P)+(p-\P)A\big)\\
+e^2(A^*A^*+AA+2A^*A)\,.
\end{multline}
$A$ depends on the cutoff $\Lambda$, which is not indicated explicitly
in our notation.

In the following we will need a smallness condition on $e$ which is
summarized as $|e|<e_0$ with suitable $e_0$ fixed throughout.
$e_0$ has its origin from several sources. It is needed for the
self-adjointness of $H_\Lambda$, for the existence of a ground
state, and in the lower bound estimate for the ground state energy
of $H_\Lambda$. In each case $e_0$ can be computed, $e_0= \Ow(1)$
in our units, but to actually carry out the integrations would not
add to the clarity of the paper.

If $|e|<e_0$, the interaction part of \eqref{6} is bounded with a
bound less than 1 relative to $H_\Lambda$ at $e=0$. At $\Lambda =
\infty$, $H_\infty$ is relatively form bounded with a bound less
than 1. Thus $H_\Lambda$, $H_\infty$ define self-adjoint operators
by the KLMN theorem \cite{RS}. We set
\begin{equation}
E^Z_\Lambda (e)= \inf \text{spec} (H_\Lambda)\,.
\end{equation}
As proved in \cite{BFS}, $E^Z_\Lambda (e)$ is an eigenvalue of
$H_\Lambda$. It persists as $\Lambda \to \infty$ \cite{HHS}. Note that
$E^Z_\Lambda (e)= E^Z_\Lambda (-e)$, since $H_\Lambda$ at $\pm e$
are unitarily equivalent.

The binding energy is the minimal energy required in ionizing the
atom. Let $T_\Lambda (p)$ denote the operator $H_\Lambda$ for
$Z=0$. $p$ appears now only as a parameter and it is known that $E
(p)= \inf \text{spec}(T_\Lambda(p))$ achieves its minimum at $p=0$
\cite{F}. Thus we define
\begin{equation}\label{9}
T_\Lambda (0) = T_\Lambda = \Hf +\P^2 + 2e (A^* \P + \P A) + e^2
(A^*A^*+ A A +2A^* A)
\end{equation}
and the self-energy
\begin{equation}
E^0_\Lambda (e)= \inf \text{spec} (T_\Lambda)\,.
\end{equation}
As $\Lambda \to \infty$, $H_\Lambda \to H_\infty$ and $T_\Lambda
\to T_\infty$ in the norm-resolvent sense. In particular, this
ensures that $\lim_{\Lambda \to \infty} E^Z_\Lambda (e)=
E^Z_\infty(e)$ and $\lim_{\Lambda \to \infty} E^0_\Lambda (e)=
E^0_\infty(e)$.
\begin{definition}\label{def1}
\it{The binding energy of $H$ is the difference}
\begin{equation}
E_{\mathrm{bin}}(e)= E^0_\infty (e) - E^Z_\infty (e)\,.
\end{equation}
It satisfies $E_{\mathrm{bin}}(e) \geq 0$.
\end{definition}

Perturbation theory means to Taylor expand $E^Z_\infty (e)$
in a power series in $e$. In such a computation one first leaves
$-Ze^2/4 \pi|x|$ unexpanded and expands only in the coupling to
the field. To lowest order one finds thereby, as to be expected,
\begin{equation}
E^{(0)}_{\mathrm{bin}}= -E_{\mathrm{at}}\,,
\end{equation}
where $E_{\mathrm{at}}$ is the ground state energy of $p^2 -
Ze^2/4 \pi|x|$, $E_{\mathrm{at}}= - \frac{1}{4}(Ze^2/4\pi)^2$. The
first radiative correction is obtained as
\begin{equation}
E^{(1)}_{\mathrm{bin}}= -E_{\mathrm{at}} \big(1 +
(e^2/6\pi^2)\big) \,.
\end{equation}
Such a computation is only formal, since 
the ground state eigenvalue of $H_\Lambda$ and of $T_\Lambda$ sits at the bottom of the continuous spectrum. The required differentiability is not
ensured through the general theory of linear operators. Still, as
our main result, we confirm the formal perturbation
theory by proving suitable upper and lower bounds.
\begin{thm}\label{thm.a}
Let $0<e<e_0$ and let $E_{\mathrm{bin}}(e)$ as given in Definition
\ref{def1}. Then there exists constants $c_+$, $c_-$, independent
of $e$, such that
\begin{equation}\label{1.14}
c_- e^7 \log (\frac{1}{e}) \leq E_{\mathrm{bin}} (e)-
(-E_{\mathrm{at}}) \big(1+ \frac{e^2}{6\pi^2}\big) \leq c_+ e^7\,.
\end{equation}
\end{thm}

\noindent
{\bf Remark}. If one reintroduces the mass
$m$ of the electron, our estimate states that
\begin{equation}\label{13}
E_{\mathrm{bin}}(e)= \frac{1}{2} m (Ze^2/4\pi)^2 \big(1+
\frac{e^2}{6\pi^2}\big) + \mathcal{O}( e^7 \log (1/e))\,.
\end{equation}
Physically, energies are calibrated in units of the effective
mass $m_\mathrm{eff}$, which is defined as the inverse curvature
of $E(p)$ at $p=0$, see above Eq. \eqref{9}. We extend in
\eqref{13} by $m_\mathrm{eff}$ and formally expand the ratio $m/
m_\mathrm{eff}$ in $e$ with the result
\begin{equation}
\frac{m}{m_\mathrm{eff}} = 1- \frac{e^2}{6\pi^2} +
\mathcal{O}(e^4)\,.
\end{equation}
Thus in \eqref{13} the relative $\mathcal{O}(e^2)$-corrections
cancel precisely. We conjecture that, after mass renormalization,
for hydrogen like atoms the radiative correction decreases the binding energy and
is of the form $E_{\mathrm{bin}}(e) = - {m_\mathrm{eff}} \eat (1 - \mathcal{O}(e^6\log(1/e))$. 
This conjecture is strongly supported by  the Lamb shift calculations, see, e.g. \cite[Eq. (5.18) and (5.23)]{HS}.
\\

The binding energy is the difference between the ground state energy
of $H_\Lambda$ and the self-energy, i.e., the ground state energy of $T_\Lambda$.
Thus to prove Theorem \ref{thm.a} we need upper and
lower bounds on $E_\Lambda^Z$, respectively $E^0_\Lambda$. In fact at order 
$e^6$ in this difference all terms except for
a single one cancel. The cases $Z=0$ and $Z \neq 0$ are handled 
by the same technique. The basic idea is to use
the perturbative ground state as a backbone. The upper bound
is easy and a straightforward application of the variational formula. The real effort
lies in the lower bound, where we employ sharp operator estimates
for the carefully corrected perturbative ground state.

The method used here was originally developed in \cite{H1}
in the context of the Pauli-Fierz Hamiltonian, including spin,  with an ultraviolet cutoff.
Catto and Hainzl \cite{CH} refined the method and extend the result
to higher order corrections. Further applications of the method are \cite{H2,HVV,HS}.

To give a brief outline: In Section \ref{Pthm1} bounds on the self-energy are established and in Section
\ref{Pthm2} we prove the corresponding bounds
for $E_\Lambda^Z$. The bounds are uniform in the cutoff $\Lambda$. Using these bounds we derive in
Section \ref{Pmt} the estimate claimed in  \eqref{1.14}. The chain of
arguments for the lower bounds is somewhat lengthy. Not to interrupt the main line
we collect all the required operator norm estimates in the Appendices \ref{Aap} - \ref{Cap}. Some of them are stated
only for completeness, while others have not been established before.

\section{Self-energy}\label{Pthm1}

In this Section we establish bounds on
$E_\Lambda^0$. It is convenient to have the shorthands
\begin{equation}
\D = \P^2 + \Hf, \qquad L= \D + 2 e^2 A^*A.
\end{equation}

\begin{thm}\label{thm1}
Let $\o$ denote the vacuum vector in $\F$. Then
\begin{multline}\label{2.2}
E^0_{\Lambda}(e) = - e^4 (\o, AA\D^{-1}A^*A^*\o) -4 e^6(\o,
AA\D^{-1}\P A\D^{-1}A^* \P \D^{-1}A^*A^*\o) \\- 4e^6 (\o,
AA\D^{-1}A^*\P\D^{-1} \P A \D^{-1}A^*A^*\o) \\+ 2e^6  (\o,
AA\D^{-1}A^*A\D^{-1}A^*A^*\o) + \Ow(e^7),
\end{multline}
where the error term $\Ow(e^7)$ is bounded uniformly in $\Lambda$.
\end{thm}
\begin{proof}
{\bf Upper bound}: In the following the expression 
$r\leq \Ow(e^m)$ means that there exists a positive constant $c$,
independent of $\Lambda$ and $e$ for $|e| \leq e_0$,
such that $r \leq c e^m$. 

We choose as  trial function the perturbative ground state
\begin{multline}\label{2.3}
\Psi =\o \oplus (-2 e^3\D^{-1} \P A \D^{-1}A^*A^*\o) \\ \oplus ( -
e^2\D^{-1}A^*A^*\o)\oplus( -2e^3
\D^{-1}A^*\P\D^{-1}A^*A^*\o).
\end{multline}
{\bf Lower bound}: As an approximate ground state (a.g.s.) we denote
any $\Psi \in \F$, such that $\|\Psi\| = 1$ and
\begin{equation}
(\Psi, T_\Lambda\Psi) \leq - e^4 (\o, AA\D^{-1}A^*A^*\o) + \Ow(e^6).
\end{equation}
Such states exist as can be seen from the upper bound. We first achieve 
an a priori bound on the
kinetic energy of $\Psi$.
\begin{lem}
Let $\Psi$ be an a.g.s., then
\begin{equation}\label{apb}
(\Psi, \D\Psi) \leq \Ow( e^4) .
\end{equation}
\end{lem}
\begin{proof}
We have
\begin{equation}
\begin{split}
(\Psi,T_\Lambda\Psi) \geq & \frac 12 (\Psi,\D \Psi) + \frac 12 (\Psi,\D \Psi) +
2\Re(\Psi,[2eA^*\P+e^2 A^*A^* ]\Psi) \\ \geq & \frac 12 (\Psi,\D \Psi) -2 \|2e
\D^{-1/2} A^*\P \Psi + e^2 \D^{-1/2} A^*A^*\Psi\|^2 \\ \geq &
\frac 12 (\Psi,\D \Psi) - \Ow (e^2)\| \P\Psi\|^2 - \Ow( e^4) \big[\|\Psi\|^2 +
\|\Hf^{1/2}\Psi\|^2\big],
\end{split}
\end{equation}
since we know from Lemma \ref{hlt1} of Appendix \ref{Cap} that $\|\D^{-1/2}A^*\| \leq c$ and $AA\D^{-1}A^*A^*
\leq c(1+\Hf)$. Here and throughout the paper $c$ will denote a generic
constant, independent of $\lambda$ and $e$.
For $e$ sufficiently small together with \eqref{apb} we arrive at the assertion of
the lemma.
\end{proof}

With our notation we can rewrite
\begin{equation}
(\Psi,T_\Lambda\Psi) = (\Psi, L\Psi) + 2 \Re (\Psi, [2eA^*\P + e^2 A^*A^*]\Psi).
\end{equation}
Observe that $L$ is invertible on $(1-P_\o)\F$. Therefore we obtain the
identity
\begin{equation}
\label{id1}
(\Psi,T_\Lambda\Psi) =- \|2eL^{-1/2}A^*\P\Psi + e^2L^{-1/2} A^*A^*\Psi\|^2 + \|L^{1/2}
h\|^2,
\end{equation}
with
\begin{equation}\label{2.9}
h= \Psi +[2eL^{-1}A^*\P + e^2L^{-1} A^*A^*]\Psi = \Psi + F^* \Psi.
\end{equation}
This notation turns out to be very convenient.
In fact this idea  allowed \cite{CH} to recover higher order corrections.
In the following we will implicitly show that $h$ is small in the sense 
$(h,\D h)\leq \Ow(e^6)$ which implies that $\Psi$ has to be close to
the perturbative ground state \eqref{2.3}.
Notice that \eqref{2.9} immediately yields $(h,\D h) \leq \Ow(e^4)$.

In the first term on the r.h.s. of \eqref{id1} we insert
$\Psi = h - F^*\Psi$, hence
\begin{equation}
\begin{split}
 (\Psi,T_\Lambda\Psi) =& - \|2eL^{-1/2}A^*\P h - 2eL^{-1/2}A^*\P F^*\Psi  \\
 &  + e^2L^{-1/2} A^*A^*\Psi\|^2 + \|L^{1/2}
h\|^2 \\=& -4e^2 \|L^{-1/2}A^*\P h\|^2 - 4e^2 \|L^{-1/2}A^*\P
F^*\Psi\|^2 \\ & - e^4 \|L^{-1/2} A^*A^*\Psi\|^2  + 2\Re \Big[ 4e^2(h,\P A
 L^{-1}A^* \P F^* \Psi) \\ & - 2e^3 (\Psi, F\P A \Li A^*A^*\Psi)+ 2e^3(h,\P A\Li
 A^*A^*\Psi)\Big] \\&+  \|L^{1/2}
h\|^2.\qquad \qquad
\end{split}
\end{equation}
First we estimate the diagonal terms. By Lemma \ref{hlt1}
\begin{equation}
e^2 \|L^{-1/2}A^*\P h\|^2 \leq e^2 \|\P h\|^2 \|\Lh A^*\|^2 \leq \Ow(e^2)  \|\P h\|^2.
\end{equation}
Slightly more care is needed for the second term,
\begin{multline}
\label{t2}
e^2\|L^{-1/2}A^*\P
F^*\Psi\|^2 = 4e^4 \|L^{-1/2}A^*\P
\Li A^*\P \Psi\|^2 \\ + 2e^5  \Re
(\Psi, \P A \Li \P A \Li A^*\P\Li A^*A^*\Psi)\\ +  e^6 \|L^{-1/2}A^*\P\Li A^*A^*\Psi\|^2 .
\end{multline}
The first term on the r.h.s. of \eqref{t2} we estimate by
\begin{equation}
e^4 \|\P \Psi\|^2 \|\Lh A^* \P \Li A^*\|^2 \leq \Ow(e^8),
\end{equation}
because of $$\|\Lh A^* \P \Li A^*\|\leq \|L^{-1/2}A^*\|\|\P
L^{-1/2}\|\|L^{-1/2}A^*\|,$$ (notice $ \|\P L^{-1/2}\|\leq \|\P
\D^{-1/2}\|\leq 1$), Lemma \ref{hlt1}, and \eqref{apb}.

Applying in a similar way Lemma \ref{hlt1} to the second term on
the r.h.s. of \eqref{t2} we obtain
\begin{multline}
e^5 |
(\Psi, \P A \Li \P A \Li A^*\P\Li A^*A^*\Psi)| \\  \leq
\Ow(e^5)\|\P\Psi\|\|(1+\Hf)^{1/2}\Psi\|\leq \Ow(e^7).
\end{multline}
Obviously,
\begin{equation}
\label{t21}
 e^6 \|L^{-1/2}A^*\P
\Li A^*A^* \Psi\|^2 \leq  e^6 \|\D^{-1/2}A^*\P\Li A^*A^* \Psi\|^2.
\end{equation}

Recall the resolvent equation
\begin{equation}\label{re}
\frac 1L = \frac 1\D - 2e^2 \frac 1\D A^*A \frac 1\D +4
e^4\frac 1\D A^*A \frac 1L A^*A \frac 1\D.
\end{equation}
Consequently, using Lemma \ref{hlt1},
\begin{multline}
 \|\D^{-1/2}A^*\P\Li A^*A^* \Psi\| \leq  \|\D^{-1/2}A^*\P\Di A^*A^* \Psi\| \\ +
2e^2 \|\D^{-1/2}A^*\P\Di A^*A\Di A^*A^* \Psi\|\\ +4 e^4  \|\D^{-1/2}A^*\P\Di
A^*A\Li A^*A \Di A^*A^*\Psi\| \\ \leq  \|\D^{-1/2}A^*\P\Di A^*A^* \Psi\| +
\Ow(e^2) \|(1 + \Hf)^{1/2}\Psi\| +\Ow(e^4) \|(1 + \Hf)^{1/2}\Psi\|.
\end{multline}
By means of Lemma \ref{she2}
\begin{multline}
(\Psi,AA\D^{-1}\P AL^{-1}A^*\P \D^{-1}A^*A^*\Psi)\\ \leq (\o, AA
\D^{-1}\P A\D^{-1}A^*\P\D^{-1} A^*A^*\o)\|\Psi\|^2 + c (\Psi,
\D\Psi).
\end{multline}
Thus we arrive at
\begin{equation}
e^2\|L^{-1/2}A^*\P F^* \Psi\|^2 \leq e^6  (\o,
AA \D^{-1}\P A\D^{-1}A^*\P\D^{-1} A^*A^*\o) + \Ow(e^8).
\end{equation}

Using again \eqref{re} we obtain
\begin{multline}
\label{hugo}
e^4\|\Lh A^*A^*\Psi\|^2 = e^4 (\Psi,AA\D^{-1}A^*A^*\Psi)\\ - 2e^6
(\Psi,AA\D^{-1}A^*A\Di A^*A^*\Psi) + 4e^8 \|\Lh A^*A\Di A^*A^*\Psi\|^2.
\end{multline}
According to Lemma \ref{she1} (cf. \cite[Lemma 1]{H1})
\begin{equation}
(\Psi,AA \D^{-1}A^*A^*\Psi) \leq (\o,
AA\D^{-1}A^*A^*\o)\|\Psi\|^2 + c(\Psi,\D\Psi)
\end{equation}
and to Lemma \ref{she3}
\begin{multline}
(\Psi,AA \D^{-1}A^*A \Di A^*A^*\Psi) \\ \geq (\o,
AA\D^{-1}A^*A \Di A^*A^*\o)\|\Psi\|^2 - c(\Psi,\D\Psi).
\end{multline}
Since 
\begin{multline}
\|\Lh A^*A\Di A^*A^*\Psi\|\leq \|\Lh A^*\|\|A\Dh\|\|\Dh
A^*A^*\Psi\|\\ \leq c \|(1+\Hf)^{-1/2}\Psi\|
\end{multline}
and using our a priori
bound \eqref{apb}, we arrive at
\begin{multline}
e^4\|\Lh A^*A^*\Psi\|^2 \leq e^4 (\o,AA \D^{-1}A^*A^*\o)\\ - 2e^6  (\o,
AA\D^{-1}A^*A \Di A^*A^*\o) + \Ow(e^8).
\end{multline}

Next we treat the off-diagonal terms.
Using Lemma \ref{hlt1} yields
\begin{multline}
e^2 |(h,\P A \Li A^* \P F^*\Psi)\| \\
\leq \Ow(e^3) \|\P h\|\|\P \Psi\| +  \Ow(e^4) \|\P h\|\|(1+\Hf)^{1/2} \Psi\| \leq
\Ow(e) \|\P h\|^2 +\Ow(e^7).
\end{multline}
By definition of $F$
\begin{multline}
\label{t3}
e^3 ( \Psi,F\P A\Li A^*A^*\Psi) \\ = 2e^4 ( \Psi,\P A \Li \P A\Li A^*A^*\Psi) +
e^5 (\Psi,AA\Li \P A\Li A^*A^*\Psi).
\end{multline}
Concerning the first term on the r.h.s., we  insert for the left vector of the
inner product  $\Psi= h - F^*\Psi$.  By Lemma \ref{hlt1}  we have
on the one hand
\begin{multline}
 e^4| ( h,\P A \Li \P A\Li A^*A^*\Psi)|\\ \leq \Ow(e^4)\|\P h\|\|(1+\Hf)^{1/2}\Psi\|
 \leq \Ow(e) \|\P h\|^2 + \Ow(e^7)
\end{multline}
and on the other hand
\begin{multline}\label{2.28}
 e^4| ( \Psi,F\P A \Li \P A\Li A^*A^*\Psi)| \leq \Ow(e^5)\|\P
 \Psi\|\|(1+\Hf)^{1/2}\Psi\| \\+
 e^6|(\Psi,AAL^{-1}\P A \Li \P A\Li A^*A^*\Psi)|.
\end{multline}
To the remaining term in \eqref{2.28} we apply the resolvent equation,
the estimates in Lemma \ref{hlt1}, as well as Lemma \ref{she4} (cf. \cite[Appendix C]{CH}), which states that 
\begin{equation}\label{appc}
|(\Psi,AA \Di \P A \Di \P A \Di A^*A^*\Psi)| \leq c \|\Psi\|\|\Hf^{1/2}\Psi\|,
\end{equation}
and
\begin{equation}
e^5| (\Psi,AA\Di \P A\Di A^*A^*\Psi)| \leq \Ow(e^5) \|\Psi\|\|\Hf^{1/2}\Psi\| .
\end{equation}
Thus we have gained
\begin{equation}
e^3 |( \Psi,F\P A\Li A^*A^*\Psi)|
 \leq \Ow(e) \|\P h\|^2 + \Ow(e^7).
\end{equation}

Assembling all together we conclude
\begin{multline}
(\Psi,T_\Lambda\Psi) \geq \\
 - e^4 (\o, AA\D^{-1}A^*A^*\o) - 4e^6(\o,
AA\D^{-1}\P A\D^{-1}A^* \P \D^{-1}A^*A^*\o) \\+ 2e^6  (\o,
AA\D^{-1}A^*A\D^{-1}A^*A^*\o) + \|L^{1/2} h\|^2 \\+ 2\Re (h, 2e^3 \P A \Li
A^*A^*\Psi) - \Ow(e) \|\P h\|^2 - \Ow(e^7).
\end{multline}
We further use the identity
\begin{multline}
\label{t4}
 \|L^{1/2} h\|^2+ 2\Re (h, 2e^3 \P A \Li
A^*A^*\Psi ) \\ =
-4\|e^3\Lh \P A \Li
A^*A^*\Psi\|^2 +  \|L^{1/2} \bar h\|^2
\end{multline}
with
\begin{equation}
\bar h = h +2e^3\Li \P A \Li A^*A^*\Psi =
h + G^* \Psi.
\end{equation}
By means of \eqref{t4} together with Lemma \ref{she2b}
we further estimate
\begin{multline}
(\Psi,T_\Lambda\Psi) \geq - e^4 (\o, AA\D^{-1}A^*A^*\o) \\- 4e^6(\o,
AA\D^{-1}\P A\D^{-1}A^* \P \D^{-1}A^*A^*\o) \\+ 2e^6  (\o,
AA\D^{-1}A^*A\D^{-1}A^*A^*\o) 
 - 4e^6 (\o,
AA\D^{-1} A^*\P\D^{-1}\P A \D^{-1}A^*A^*\o)\\ 
- \Ow(e^6) (\Psi, \D\Psi) +
\|L^{1/2} \bar h\|^2- \Ow(e) \|\P \bar h\|^2 - \Ow(e) \|G^*\Psi\|^2 -
\Ow(e^7) \\ \geq
- e^4 (\o, AA\D^{-1}A^*A^*\o) - 4e^6(\o,
AA\D^{-1}\P A\D^{-1}A^* \P \D^{-1}A^*A^*\o) \\+ 2e^6  (\o,
AA\D^{-1}A^*A\D^{-1}A^*A^*\o) \\- 4e^6 (\o,
AA\D^{-1} A^*\P\D^{-1}\P A \D^{-1}A^*A^*\o) 
- \Ow(e^7),
\end{multline}
for $e$ small enough such that
\begin{equation}
\|L^{1/2} \bar h\|^2- \Ow(e) \|\P \bar h\|^2 \geq \|\P \bar h\|^2 (1-  \Ow(e)
) \geq 0,
\end{equation}
which concludes the proof of Theorem \ref{thm1}.
\end{proof}

\section{Ground state energy}\label{Pthm2}

In the following we denote $ V = - \frac {e^2}{4\pi |x|}$ and $\phi= \phi_0
\otimes \o$, where $\phi_0$ is the ground state of the Schr\"odinger operator
$p^2 + V$, with ground state energy $\eat$, i.e.,
\begin{equation}
(p^2 + V)\phi_0 =\eat \phi_0.
\end{equation}
Since $\eat= -\frac 1{4(4\pi)^2}Z^2 e^4$, we observe, e.g., by virial theorem, 
$\|p\phi\|^2 = \Ow(e^4)$.
For convenience we introduce the notation
\begin{eqnarray*}
P &=& p - \P,\\
B &=& P^2 + V - \eat + \Hf ,\\
K &=& B + 2 e^2 A^*A.\\
\end{eqnarray*}

\begin{thm}\label{thm2}
\begin{multline}\label{gse}
E^Z_\Lambda(e) = \eat - 4 e^2 (\phi,pA B^{-1}A^*p\phi) - e^4
(\phi,AAB^{-1}A^*A^*\phi)\\ - 4 e^6(\phi,
AAB^{-1}\P AB^{-1}A^* \P B^{-1}A^*A^*\phi)\\
- 4 e^6 (\phi,AAB^{-1}A^* \P B^{-1}\P A B^{-1}A^*A^*\phi) \\+ 2 e^6 (\phi,
AAB^{-1} A^*AB^{-1}A^*A^*\phi) + \Ow\big(e^7 \ln(1/e)\big)
\end{multline}
uniformly in $\Lambda$.
\end{thm}

\subsection{Upper bound}
As in Section \ref{Pthm1} we use the perturbative ground state
\begin{multline}
 \Psi = \phi\oplus ( -2eB^{-1}A^*p\phi - 2e^3 B^{-1}P A  B^{-1}A^*A^*
\phi)\\
\oplus(4e^2B^{-1}A^*P B^{-1}A^*p\phi - 2e^2 B^{-1} A^*A^*\phi)\oplus( -2e^3B^{-1}A^*P  B^{-1} A^*A^*
\phi)
\end{multline}
Apart from error terms which are at least of order $\Ow(e^7)$ we obtain
\begin{multline}\label{45}
( \Psi, H \Psi) = \eat - 4 e^2(\phi,p AB^{-1}A^* p \phi) -  e^4 (\phi,AAB^{-1}A^*A^*\phi) \\ +2 e^6
(\phi,AAB^{-1}A^*AB^{-1}A^*A^*\phi)- 4 e^6
(\phi,AAB^{-1}P AB^{-1}A^*P B^{-1} A^*A^*\phi)\\ -
4e^6 (\phi, AAB^{-1}A^* P
B^{-1}P AB^{-1}A^*A^* \phi)  \\ +8e^4 \Re(\phi,p AB^{-1}P AB^{-1}A^*A^* \phi)  +
\Ow(e^7).
\end{multline}
The last three terms can be simplified further by taking advantage of the fact that
we deal with the Coulomb potential $V= -\frac{e^2}{4\pi|x|}$.
Namely we insert $P=p - \P$ and show that
all terms resulting from the summand  $p$ are of higher order.
For this purpose we transform canonically as  
\begin{equation}
x\to x/e^2, \,\,\, p\to e^2 p,
\end{equation}
i.e., through the unitary $U_e$ as
\begin{equation}
U_e p^2 U_e^*= e^4 p^2, \,\,\, U_e( p^2 - \frac{e^2}{4\pi|x|}) U_e^*= e^4 (p^2 - \frac{1}{4\pi|x|}).
\end{equation}
By means of that
transformation we estimate, e.g.,
\begin{multline}
(\phi,AAB^{-1}p AB^{-1}A^*p B^{-1} A^*A^*\phi)=\\(U_e\phi,U_e AAB^{-1}p AB^{-1}A^*p B^{-1} A^*A^*U_e^*U_e\phi) \leq
\\ \|\phi\|^2 \|B_e^{-1/2}A^* e^2 p^2B_e^{-1} A^*A^*\|^2 \leq \Ow(e^4),
\end{multline}
where $B_e = (e^2 p - \P)^2 + e^2 V -\eat + \Hf$.
Using  similar estimates, together with Schwarz inequality,
we see that in the last three terms in \eqref{45}, apart from higher order
terms, only expressions involving $\P$ play a role.

Finally notice that
\begin{multline}
|(\phi,p AB^{-1}P AB^{-1}A^*A^* \phi)| = e^2 |(U_e\phi, p AB_e^{-1} P AB_e^{-1}A^*A^* U_e \phi)|\\
\leq \Ow(e^4),
\end{multline}
which follows from expanding $1/B_e$ and the fact that
the lowest order term vanishes, since
$(U_e\phi, p U_e\phi)=0$.
\subsection{Lower bound}
We recall the convention on $\Ow(e^m)$ from Section \ref{Pthm1}.

The proof of the lower bound proceeds in analogy  to
Theorem \ref{thm1}. The decisive difference is that
 we deal now with operators $B$, $K$, and $P$ which do not commute which
means that we have to be a bit more carefully in our
estimates. Apart from that the strategy is not altered.

As in Section \ref{Pthm1} we consider an a.g.s. satisfying $(\Psi,(H-\eat)\Psi) \leq \Ow(e^4)$.
Because $\eat = \Ow(e^4)$ we conclude the bound
\begin{equation}\label{ap}
(\Psi, [P^2 + \Hf]\Psi) \leq \Ow(e^4).
\end{equation}
Note that the existence of a true ground state is not needed for the argument.
We can write
\begin{multline}\label{Hbound}
(\Psi,H\Psi) =  \eat\|\Psi\|^2 +  (\Psi,K\Psi) + 2\Re(\Psi,2eA^*P\Psi + e^2 A^*A^*\Psi).
\end{multline}
Following the same scheme as in \eqref{id1}  we obtain the identity
\begin{multline}\label{epsi2}
(\Psi,H\Psi)=  \eat\|\Psi\|^2  -\|
2eK^{-1/2}A^*P \Psi + e^2K^{-1/2} A^*A^* \Psi\|^2 +
\|K^{1/2} h\|^2,
\end{multline}
with
\begin{equation}\label{3.12}
h = \Psi + 2e K^{-1} A^*P \Psi + e^2 K^{-1}
A^*A^* \Psi = \Psi + F^*\Psi.
\end{equation}
Notice that $h$ also fulfills $(h, [P^2 + \Hf]h) \leq \Ow(e^4)$.

Some of the terms in the lower bound 
are logarithmically infrared divergent. In this case 
we replace $\Hf$ by $\Hf + e^7$, which causes the additional 
error 
$-e^7 \|\Psi\|^2$ in the r.h.s. of
(\ref{Hbound}). Also the bound acquires a logarithmic correction.

We  now insert 
\begin{equation}
\Psi=h -2eK^{-1}A^*P\Psi- e^2 K^{-1}A^*A^*\Psi,
\end{equation}
in (\ref{epsi2}) in order to obtain
\begin{multline}\label{eq1}
(\Psi,H \Psi) = \eat - \| -4e^2K^{-1/2}
A^*P K^{-1} A^*P \Psi \\- 2e^3 K^{-1/2} A^*PK^{-1}
A^*A^*\Psi + 2e K^{-1/2}A^* P h \\+ e^2 K^{-1/2}
A^*A^*\Psi\|^2 + \|K^{1/2}h\|^2,
\end{multline}
 recall that $\|\Psi\|=1$.
Multiplying out the norm we observe that  (\ref{eq1}) is equal to
\begin{eqnarray}\label{sum1}
 \eat + \|K^{1/2} h\|^2 & -& 4e^2(h,P AK^{-1}A^*P h)\\
\label{sum2} & -&e^4(\Psi,AAK^{-1}A^*A^*\Psi)\\
\label{sum3} &
-&16e^4(\Psi,P AK^{-1}P AK^{-1}A^*PK^{-1}A^*P\Psi)\\
\label{sum4} &-&4e^6(\Psi,AAK^{-1} P AK^{-1}A^*PK^{-1}A^*A^*\Psi)\\
\label{sum5} &+& 2\Re
\Big[4e^4(\Psi,P AK^{-1}P AK^{-1}A^*A^*\Psi)\\
\label{sum6} & +& 4e^4(\Psi,AAK^{-1}P AK^{-1}A^*P
h)\\
\label{sum7} & +&
2e^5(\Psi,AAK^{-1}P AK^{-1}A^*A^*\Psi)\\ \label{sum8}&
-& 2e^3 (h, P AK^{-1}A^*A^*\Psi)\\
\label{sum9} &-& 8 e^5 (\Psi,P
AK^{-1}P AK^{-1}A^*PK^{-1}A^*A^*\Psi)\\ \label{sum10}&
-& 8e^3 (\Psi, P AK^{-1}P AK^{-1}A^*P h)\Big].
\end{eqnarray}
Applying Lemma \ref{hlt3} $(iii)$ together with \eqref{ap} we immediately obtain
\begin{equation}
|\eqref{sum3}| \leq \Ow(e^4\ln(1/e)) \|P \Psi\|^2 \leq \Ow(e^8\ln(1/e))
\end{equation}
and by  Lemma \ref{hlt3} $(iii)$ and $(iv)$, the bounds
$$| (\ref{sum9})|\leq  \Ow(e^5\ln(1/e)) \|P\Psi\|\|(1+\Hf)^{1/2}\Psi\| \leq \Ow(e^7\ln(1/e)),$$
respectively, by Lemma \ref{hlt2} $(i)$ and Lemma \ref{hlt3} $(iii)$
 $$| (\ref{sum10})| \leq \Ow(e^3)
\big(\|P\Psi\|\|P h\| \big) \leq \Ow(e^7).$$
Additionally by Lemma \ref{alem5} we can bound
\eqref{sum7} by
\begin{multline}
|\eqref{sum7}|= 2e^5 | (\Psi,AA K^{-1}P AK^{-1} A^*A^* \Psi)|\\ \leq \Ow(e^5\ln(1/e))\|\Psi\| \|\Hf^{1/2}\Psi\| \leq \Ow(e^7\ln(1/e)).
\end{multline}
In the remaining terms, apart from \eqref{sum6} and \eqref{sum8}, we insert again
$$\Psi = h - 2eK^{-1}A^*P \Psi - e^2 K^{-1}
A^*A^*\Psi.$$

Applying our inequalities in Lemma \ref{hlt2} and \ref{hlt3}  we infer
\begin{equation}\label{432}
\begin{split}
(\Psi,H\Psi)
\geq  & - 4e^2(h,P AK^{-1}A^*P h)\\
& -e^4(h,AAK^{-1}A^*A^*h)\\
 &-4e^6(h,AAK^{-1} P AK^{-1}A^*PK^{-1}A^*A^*h)\\
&+2
e^4 \Re(h,4P AK^{-1}P AK^{-1}A^*A^*h)\\
&-2 e^3 \Re(h,2 P AK^{-1}A^*A^*\Psi)\\
&+2 e^4 \Re(h,4 PAK^{-1}A^*PK^{-1}A^*A^*\Psi)
\\&+ \|K^{1/2} h\|^2 + \eat  -\Ow(e^7\ln(1/e)).
\end{split}
\end{equation}
Neglecting first the terms $ \eat -  \Ow(e^7\ln(1/e))$ we rewrite \eqref{432} in the shorthand
\begin{equation}\label{hhh}
(h,[K + \dv]h) - 2 \Re (h,[2e^3 P AK^{-1}A^*A^* - 4 e^4 PAK^{-1}A^*PK^{-1}A^*A^*]  \Psi ),
\end{equation}
where
\begin{multline}
\dv= -4e^2P AK^{-1}A^*P
 -e^4 AAK^{-1}A^*A^* \\
+8e^4\Re[ P AK^{-1}P AK^{-1}A^*A^*]
 -4e^6 AAK^{-1} P AK^{-1}A^*PK^{-1}A^*A^*.
\end{multline}
Since, due to the  Lemmas in Appendix  \ref{Bap}, $R$ is relatively bounded to $K$, we conclude
that for $e$ small enough
\begin{equation}\label{relb}
K + \dv \geq - C e^4 :=- \mu.
\end{equation}
In fact, by Lemma \ref{alem2} to \ref{alem4}, and Lemma \ref{hlt3} we see that
for $e$ small enough, i.e., for those $e$ such that $\Hf$ in $K$ dominates the error
terms from $R$, $$K + \dv \geq (1 -c e^2) P^2 + V - c' e^4 \geq -  C e^4,$$
for appropriate constants which implies \eqref{relb}.

Therefore (\ref{hhh}) can be rewritten as
\begin{multline}\label{fff}
 (\ref{hhh})=- \mu\|h\|^2 + (h,[K + \dv + \mu]h)\\ - 2 \Re (h,[2e^3 P AK^{-1}A^*A^* -4 PAK^{-1}A^*PK^{-1}A^*A^*]  \Psi ) \\
= - \mu\|h\|^2+ \|[K
+\dv+\mu]^{1/2}\bar h\|^2\\- 4e^6\|[K + \dv+\mu]^{-1/2}[P AK^{-1}A^*A^*  - 2ePAK^{-1}A^*PK^{-1}A^*A^*]\Psi\|^2 ,
\end{multline}
with
\begin{multline}\label{defhb}
\bar h = h +2 e^3[K + \dv+\mu]^{-1}P A K^{-1}A^*A^*\Psi \\ -  4 e^4[K + \dv+\mu]^{-1} PAK^{-1}A^*PK^{-1}A^*A^*\Psi.
\end{multline}
Using \eqref{defhb}, Lemma \ref{hlt2} and \ref{hlt3}, together with the fact that $\mu$ is of order $e^4$ we estimate
\begin{multline}
\eqref{fff} =   - \mu\|h\|^2+ \|[K
+\dv+\mu]^{1/2} \bar h\|^2 \\- 4e^6\|[K + \dv+\mu]^{-1/2}[P AK^{-1}A^*A^*  - 2ePAK^{-1}A^*PK^{-1}A^*A^*]\Psi\|^2
\\\geq  \|[K
+\dv]^{1/2} \bar h\|^2 -4e^6\|[K + \dv+\mu]^{-1/2}P AK^{-1}A^*A^*\Psi\|^2 - \Ow(e^7\ln(1/e)).
\end{multline}

Apart from errors of order $\Ow(e^7)$ we can set $\Psi = \bar
h$ and $\bar h = h$. Consequently,
\begin{multline}
(\Psi, H\Psi) \geq  \eat \\ +
(\bar h, [K + \dv] \bar h) - 4e^6 (\bar h, AAK^{-1}A^* P
K^{-1}P AK^{-1}A^*A^* \bar h)\\ - \Ow(e^7\ln(1/e)),
\end{multline}
which leads to
\begin{multline}
(\Psi, H\Psi) \geq \eat  \\+(\bar h,K \bar h) - \Big(4e^2(\bar h,P AK^{-1}A^*P \bar h) +  e^4 (\bar h,AAK^{-1}A^*A^*\bar h)
\\+  4e^6
(\bar h,AAK^{-1}P AK^{-1}A^*PK^{-1} A^*A^* \bar h)\\ +
4e^6 (\bar h, AAK^{-1}A^* P
K^{-1}P AK^{-1}A^*A^* \bar h)\\ - 8e^4 \Re(\bar h,P AK^{-1}P AK^{-1}A^*A^* \bar h) \Big)
-\Ow(e^7\ln(1/e)).
\end{multline}

Next, we extract the $e^2 2 A^*A$-term. To this aim recall
 $K = B + e^2 2 A^*A$, use the resolvent equation  (\ref{re}), the
 operator inequalities in Lemma \ref{hlt2}, and Lemma \ref{hlt3}.
We obtain
\begin{multline}\label{ende}
(\Psi, H\Psi) \geq  \eat  \\+ (\bar h,B \bar h) - \Big(4e^2(\bar h,P AB^{-1}A^*P \bar h) +  e^4 (\bar h,AAB^{-1}A^*A^*\bar h) \\
- 2e^6
(\bar h,AAB^{-1}A^*AB^{-1}A^*A^*\bar h)+  4e^6
(\bar h,AAB^{-1}P AB^{-1}A^*P B^{-1} A^*A^*\bar h)\\ +
4 e^6 (\bar h, AAB^{-1}A^* P
B^{-1}P AB^{-1}A^*A^* \bar h)  \\ -8 e^4 \Re(\bar h,P AB^{-1}P AB^{-1}A^*A^* \bar h)\Big)
-\Ow(e^7\ln(1/e)).
\end{multline}
To the terms inside the bracket we apply now
 Lemma \ref{alem2} to \ref{alem4}. 
The error terms corresponding to these Lemmas 
are bounded from below by $$-\Ow(e^4\ln(1/e)) \big(\|P\bar h\|^2 +(\bar h, \Hf\bar h)\big).$$
Recall $B= P^2 + V -\eat + \Hf$. Therefore the error  $- \Ow(e^4\ln(1/e))(\bar h, \Hf\bar h)$
is controlled by $(\bar h, \Hf\bar h)$ for $e$ small enough.
Since $\|P\bar h\|^2 \leq \Ow(e^4)$, we infer
\begin{multline}\label{ende1}
(\Psi, H\Psi) \geq  \eat  + (\bar h,[P^2 + V - \eat] \bar h) \\
- \Big(4e^2(\bar h,P AB^{-1}A^*P \bar h) +  e^4 (\bar h \vp\c\vp,B^{-1}\bar h\vp\c\vp) \\
- 2e^6
(\bar h \vp\c\vp B^{-1}A^*AB^{-1}\bar h \vp\c\vp)+  4e^6
(\bar h\vp\c \vp,B^{-1}P AB^{-1}A^*P B^{-1} \bar h\vp \c \vp)\\ +
4 e^6 (\bar h\vp\c \vp ,B^{-1}A^* P
B^{-1}P AB^{-1} \bar h \vp\c\vp)   \\ -8 e^4 \Re(\vp  \c P B^{-1} \vp\c P\bar h, B^{-1}  \bar h \vp\c \vp)\Big)
-\Ow(e^7\ln(1/e))
\end{multline}
where we used the notation
 
$$ [\bar h\vp\c \vp]_{n+2} = \bar h_n(x,k_1,\dots,k_n) \vp(k_{n+1})\c \vp(k_{n+2})$$
as introduced in \eqref{B6}.
By Lemma \ref{hlt2} (i) and Lemma \ref{alem3}
the first and the last term in the bracket are bounded by $(e^2 + e^4) \|P\bar h\|^2 + e^4 \|\bar h\|^2$.
Therefore they are relatively bounded with respect to  $ P^2 + V -\eat$.
Due to Lemmas \ref{alem2} to \ref{alem3b} the other terms in the bracket
are bounded. 
Since $ P^2 + V -\eat$ has $0$ as isolated eigenvalue, 
we are now in the favorable position to apply Kato's perturbation theory \cite{K}.

To illustrate, for simplicity, we concentrate on one term
inside the bracket, e.g., the term corresponding to $ e^4 (\bar
h\vp\c \vp,B^{-1}\bar h \vp\c\vp) $.
In other words we search for the
ground state energy of
\begin{equation}\label{kat}
(\bar h,[ P^2 + V -\eat ]\bar h) - e^4 (\bar
h\vp\c \vp,B^{-1}\bar h \vp\c\vp).
\end{equation}
Recall $\phi$ is the unique ground state of  $ P^2 + V -\eat$ with eigenvalue
$0$
therefore due to Kato
\begin{equation}
\eqref{kat} = -e^4 (\phi, AA B^{-1} A^*A^*\phi)\|\bar h\|^2 + \Ow(e^8)\|\bar h\|^2,
\end{equation}
since
\begin{equation}
(\phi \vp\c\vp,B^{-1} \phi \vp\c\vp) = (\phi, AA B^{-1} A^*A^*\phi).
\end{equation}
Remark that from Lemma \ref{hlt2} and \eqref{ap} together with
definition \eqref{3.12} and \eqref{defhb} we obtain
$$ |\|\bar h\|^2 - 1| \leq \Ow(e^3\ln(1/e)).$$
Consequently
\begin{equation}
\eqref{kat} = -e^4 (\phi, AA B^{-1} A^*A^*\phi) + \Ow(e^7\ln(1/e)).
\end{equation}
Using this strategy for each term in the bracket of \eqref{ende1}
and noticing $\|P \phi\|^2 = \Ow(e^4)$, we  obtain
an equation equivalent to \eqref{45}, this time with an error
of order $\Ow(e^7\ln(1/e))$.
Finally we  use  the considerations from the upper bound and
conclude the proof of Theorem \ref{thm2}.

\section{Proof of Theorem \ref{thm.a}}\label{Pmt}

To complete the proof of Theorem \ref{thm.a} we only have to 
work out the leading terms in \eqref{2.2} and \eqref{gse} and to
show that the difference agrees with \eqref{1.14} up to errors 
of order $e^7$. For this purpose we use the resolvent expansion
\begin{equation}\label{form}
\frac 1B= \frac 1Q -  \frac 1Q b\frac 1Q +  \frac 1Q b\frac 1B b \frac 1Q
\end{equation}
with $Q = p^2 +V - \eat + \Hf +\P^2$ and $b= - 2p
\P$. \eqref{form} is inserted in \eqref{gse}. The terms linear in $p$ vanish and the quadratic terms
are of order $\Ow(e^8)$, since 
$(\phi_0,p^2\phi_0) = - 2 \eat$. Thus only the term $Q^{-1}$ remains.
Comparing it with \eqref{2.2} we note that all terms in Eq.  \eqref{2.2} are
canceled. The only 
contribution remaining is then 
\begin{equation}
\label{4.2} 
E_\Lambda^0 - E_\Lambda^Z = - \eat
+ 4 e^2 (\phi, pA Q^{-1} A^* p \phi) + \Ow(e^7 \log e).
\end{equation}
The scalar product in \eqref{4.2} reads, to lowest order,
\begin{equation}
-\eat \frac 43 e^2(2\pi)^{-3} \int_{|k|\leq \Lambda} dk \frac 1{2|k|} k^2 (|k| + k^2)^{-3}.
\end{equation}
Taking the limit $\Lambda \to \infty$, using that all error bounds are uniform in
$\Lambda$, proves \eqref{1.14}.
\begin{appendix}

\section{Sharp estimates needed for Theorem \ref{thm1}}
\label{Aap}

We collect sharp inequalities
as used in the proof of Theorem \ref{thm1} and 
proceed analogously to \cite{H1,CH},
with the slight difference that
we have to take care of the uniform boundedness
of the error terms in the cutoff $\Lambda$.

For this aim notice that
for $s \in (0,1)$
\begin{equation}\label{defs}
\int dk \left| \frac{|\varphi(k)|}{|k|^{s}}\right|^2 \leq \frac{C}{s(1-s)},
\end{equation}
where the constant $C$ is independent of the cutoff.
For later purposes we also define
\begin{equation}
c_I =  \int \frac{|\varphi(k)|^2}{|k|^{1/2}} dk, \qquad
c_{II} =  \int \frac{|\varphi(k)|^2}{|k|}dk.
\end{equation}
Recall that
\begin{equation*}
\D = \P^2 + \Hf.
\end{equation*}

\begin{lem}\label{she1}
\begin{equation}
(\Psi,AA \Di A^*A^*\Psi) \leq (\o,AA\Di A^*A^*\o) \|\Psi\|^2 + c (\Psi,\D \Psi)
\end{equation}
with $c$ uniformly bounded in $\Lambda$.
\end{lem}
\begin{proof}
The proof follows \cite[Lemma 1]{H1}.
Fix  the photon number $n$ and recall
\begin{multline}\label{defDD}
[A^* A^*\psi_n]_{n+2} = \frac 1{\sqrt{(n+2)(n+1)}}
 \sum_{j=1}^{n+2} \sum_{\substack{i=1 \\ i \neq j}}^{n+2}
\varphi(k_j)\cdot \varphi(k_i) \times\\\times
\psi_n(k_1,\dots,\not\!\! k_j,\dots,\not \!\! k_i,\dots,k_{n+2}),
\end{multline}
where $\not \!\! k_j$ indicates that the $j-$th variable is
omitted. Using permutation symmetry we distinguish between
three different terms,
\begin{equation}\label{defI}
\big( \psi_n,AA \Di  A^* A^* \psi_n \big)=
I_n+II_n+III_n,
\end{equation}
which result naturally once we insert Equation (\ref{defDD}) into
(\ref{defI}) and
have in mind that the l.h.s. of (\ref{defI}) can be written as
\begin{equation}\label{defDD2}
\big( A^* A^* \psi_n,  [\P^2 + \Hf]^{-1} A^*A^* \psi_n \big).
\end{equation}
The most important diagonal term reads
\begin{equation}\label{A7}
I_n=  2 \int \frac{\big[\varphi(k_1)\cdot
\varphi(k_2)\big]^2
|\psi_n(k_3,\dots,k_{n+2})|^2}{\big|\sum_{i=1}^{n+2}k_i\big|^2 +
\sum_{i=1}^{n+2}|k_i|}  dk_1\dots dk_{n+2}.
\end{equation}
If we set $Q =\big|\sum_{i=3}^{n+2}k_i\big|^2 + \big|k_1 +
k_2\big|^2+ \sum_{i=1}^{n+2}|k_i|$ and $b= -
2\big[\sum_{i=3}^{n+2}k_i\big]\cdot \big[k_1+k_2\big]$ and use the
expansion
\eqref{form}
then we see that the second term vanishes when integrating over
$k_1,k_2$. Therefore, with $Q \geq \big|k_1 + k_2\big|^2+ |k_1| +
|k_2|$ and $Q+b \geq |k_1| + |k_2|$ we arrive at
\begin{multline}
I_n\leq 2 \Big[ \|\psi_n\|^2\int
\frac{\big[\varphi(k_1)\cdot\varphi(k_2)\big]^2}{|k_1 + k_2|^2 +
|k_1| + |k_2|} dk_1 dk_2 \\ + 4\int
\frac{\big|\varphi(k_1)\big|^2\big|\varphi(k_2)\big|^2 \big[|k_1|
+ |k_2|\big]^2}{\big[|k_1 + k_2|^2 + |k_1| +
|k_2|\big]^2(|k_1|+|k_2|)}\times \\ \times
\big|\sum_{i=3}^{n+2}k_i\big|^2|\psi_n(k_3,\dots,k_{n+2})|^2dk_1\dots
dk_{n+2}\Big] \\ \leq (\o,AA \Di A^*A^*\o)
\|\psi_n\|^2 + c_I^2  \|\P\psi_n\|^2.
\end{multline}

Furthermore, by use of Schwarz inequality,
\begin{multline}\label{she1e2}
II_n \leq n
\int\frac{\big|\varphi(k_1)\big| \big|
\varphi(k_2)\big|\big|\varphi(k_1)\big|\big|
\varphi(k_{n+2})\big|}{\sum_{i=1}^{n+2} |k_i|} \times \\ \times
|\psi_n(k_3,\dots,k_{n+2})||\psi_n(k_2,\dots,k_{n+1})| dk_1\dots
dk_{n+2}\\ \leq
\int \frac{|\varphi(k_1)|^2}{|k_1|}dk_1 \Big(\frac{|\varphi(k_2)|}{|k_2|^{1/2}}|k_{n+2}|^{1/2}
|\psi_n(k_3,\dots,k_{n+2})|,\times \\ \times
\frac{|\varphi(k_{n+2})|}{|k_{n+2}|^{1/2}}|k_{2}|^{1/2}|\psi_n(k_2,\dots,k_{n+1})|\Big)
\\ \leq c_{II}^2 (\psi_n,\Hf \psi_n).
\end{multline}

For the third term we  use Schwarz again
to obtain
\begin{multline}
III_n \leq n^2
\int\frac{\big|\varphi(k_1)\big|\big|
\varphi(k_2)\big|\big|\varphi(k_{n+1})\big| \big|
\varphi(k_{n+2})\big|}{\sum_{i=1}^{n+2} |k_i|} \times \\ \times
|\psi_n(k_3,\dots,k_{n+2})||\psi_n(k_1,\dots,k_{n})| dk_1\dots
dk_{n+2}\\ \leq n^2\Big(\frac{ |\varphi(k_1)||
\varphi(k_2)|}{|k_1|^{1/2}|k_2|^{1/2}}|k_{n+1}|^{1/2}|k_{n+2}|^{1/2}|\psi_n(k_3,\dots,k_{n+2})|,
\frac 1{\Hf} \times \\ \times|k_1|^{1/2}|k_2|^{1/2} |\psi_n(k_1,\dots,k_{n})|
\frac{|\varphi(k_{n+1})||\varphi(k_{n+2})|}{|k_{n+1}|^{1/2}|k_{n+2}|^{1/2}}\Big) \\ \leq c_{II}^2 n \int
|k_{n+2}|\sum_{i=3}^{n+1}
|k_i|\frac{|\psi_n(k_3,\dots,k_{n+2})|^2}{\sum_{i=3}^{n+1}
|k_i|}dk_3...dk_{n+2} \\ \leq c_{II}^2 n \int |k_{n+2} ||\psi_n|^2 dk_3...dk_{n+2} 
= c_{II}^2 (\psi_n, \Hf \psi_n).
\end{multline}
By summing over the photon number $n$ we arrive at  the statement of the Lemma.
\end{proof}

All following lemmas are proven by a scheme similar to Lemma \ref{she1}.
To shorten the calculations we introduce the operator $|A|$, which is defined by
replacing $\varphi$ in $A$ by $|\varphi|$, i.e.,
\begin{equation}
|A| = \int |\varphi(k)|a(k)dk.
\end{equation}
$|A|^*$ denotes its operator adjoint.
In essence by \eqref{she1e2} one has
\begin{equation}
|A|^*|A| \leq c_A \Hf
\end{equation}
with $c_A= \int\frac{|\varphi(k)|^2}{|k|}dk$.
Similar methods were used in \cite{H1,CH}.
In addition, in order to simplify the notation,
we introduce
\begin{equation}
P_j^l = \sum_{i=j}^l k_i, \quad H_j^l = \sum_{i=j}^l |k_i|.
\end{equation}

\begin{lem}\label{she2}
\begin{multline}
(\Psi,AA \Di \P A \Di A^* \P\Di A^*A^*\Psi) \\
\leq (\o,AA \Di \P A \Di A^* \P\Di A^*A^*\o) \|\Psi\|^2 + c (\Psi,\D \Psi)
\end{multline}
with $c$ uniformly bounded in $\Lambda$.
\end{lem}
\begin{proof}
Following the scheme of Lemma \ref{she1}
we can now distinguish between four different terms,
since there are three photons created.

The diagonal and most interesting part reads
\begin{multline}
I_n= \int \frac{[\varphi(k_{n+1})\c\varphi(k_{n+2}) ]^2 [\varphi(k_{n+3})\c P_1^{n+2}]^2 |\psi_n(k_1,\dots,k_n)|^2}{
[( P_1^{n+2})^2 + H_1^{n+2}]^2 [( P_1^{n+3})^2 + H_1^{n+3}]}dk_1\dots dk_{n+3} \\ \leq
 \int \frac{[\varphi(k_{n+1})\c\varphi(k_{n+2}) ]^2 [\varphi(k_{n+3})\c P_1^{n+2}]^2 |\psi_n(k_1,\dots,k_n)|^2
 dk_1\dots dk_{n+3} }{
[( P_{n+1}^{n+2})^2 + H_{n+1}^{n+2} + 2 P_1^n \c P_{n+1}^{n+2} ]^2 [( P_{n+1}^{n+3})^2 +  H_{n+1}^{n+3}
+ 2 P_1^n \c P_{n+1}^{n+3}]}.
\end{multline}
In order to expand the denominator we write
\begin{multline}\label{A16}
\frac 1{(Q_1 + b_1)^2 (Q_2 + b_2)} \\ =
\left[ \frac 1{Q_1^2} - \frac{2b_1}{Q_1^2 (Q_1 + b_1)} + \frac{b_1^2}{Q_1^2 (Q_1 + b_1)^2}\right] \left[
\frac 1{Q_2} - \frac{b_2}{Q_2 (Q_2 + b_2)}\right] \\ =  \frac 1{Q_1^2Q_2} +
M
\end{multline}
with $Q_1 =( P_{n+1}^{n+2})^2 + H_{n+1}^{n+2}$, $b_1 =2 P_1^n \c
P_{n+1}^{n+2}$ and the equivalent expression for $Q_2,b_2$.

The most important term is the one involving
$\frac 1{Q_1^2Q_2}$, i.e.,
\begin{multline}
 \int \frac{[\varphi(k_{n+1})\c\varphi(k_{n+2}) ]^2 [\varphi(k_{n+3})\c( P_{n+1}^{n+2}+ P_1^{n})]^2
 |\psi_n(k_1,\dots,k_n)|^2}{
[( P_{n+1}^{n+2})^2 + H_{n+1}^{n+2} ]^2 [( P_{n+1}^{n+3})^2 +  H_{n+1}^{n+3}
]}dk_1\dots dk_{n+3} \\
\leq (\o,AA \Di \P A \Di A^* \P\Di A^*A^*\o) \|\psi_n\|^2 \\ +
\int  \frac{|\varphi(k_{1})|^2|\varphi(k_{2})|^2| \varphi(k_{3})|^2}{ (H_{1}^{2})^2
H_{1}^{3}}dk_{1}dk_2dk_{3}\,
(\psi_n,\P^2\psi_n)\\
+ 2 \int  \frac{|\varphi(k_{1})|^2|\varphi(k_{2})|^2| \varphi(k_{3})|^2}{ (H_{1}^{2})^{1/2} H_{1}^{3}}dk_{1}dk_2dk_{3}
(\psi_n,\Hf\psi_n),
\end{multline}
where we used $$\frac{P_{n+1}^{n+2} }{(P_{n+1}^{n+2})^2 + H_{n+1}^{n+2} } \leq \frac 1{2
(H_{n+1}^{n+2})^{1/2}}$$ and then changed variables to simplify the
notation.
Observe $$\int  \frac{|\varphi(k_{1})|^2|\varphi(k_{2})|^2|| \varphi(k_{3})|^2}{ (H_{1}^{2})^2 H_{1}^{3}}dk_{1}dk_2dk_{3}
\leq c_{II}^3$$
and
$$  \int  \frac{|\varphi(k_{1})|^2|\varphi(k_{2})|^2| \varphi(k_{3})|^2}{ (H_{1}^{2})^{1/2} H_{1}^{3}}dk_{1}dk_2dk_{3}
\leq \int \frac{|\varphi(k_{1})|^2|\varphi(k_{2})|^2| \varphi(k_{3})|^2}{|k_1|^{1/4}|k_2|^{1/4}|k_3|}dk_{1}dk_2dk_{3}$$
which are obviously uniformly bounded.

Estimating the terms involving $M$ works similar to the two last terms.
It is a straightforward but lengthy calculation, hence skipped.

In $II_n$, where only one index differs, we meet the term
\begin{multline}
n \int \frac{[\varphi(k_{n+1})\c\varphi(k_{n+2}) ]^2 |\varphi(k_{n+3})\c P_1^{n+2}| |\varphi(k_{1})\c P_2^{n+3}|\times}{
[( P_1^{n+2})^2 + H_1^{n+2}] [( P_2^{n+3})^2 + H_2^{n+3}][( P_1^{n+3})^2 + H_1^{n+3}]}\\ \times |\psi_n(k_1,\dots,k_n)|
|\psi_n(k_2,\dots,k_n,k_{n+3})|dk_1\dots dk_{n+3} \\ \leq c_I^2 (|\psi_n|,|A|^*|A||\psi_n|)\leq c_I^2c_A (\psi_n,\Hf \psi_n)
\end{multline}
and the term
\begin{multline}
n \int \frac{|\varphi(k_{n+1})||\varphi(k_{n+2})||\varphi(k_1)|  [\varphi(k_{n+3})\c P_1^{n+2}]^2\times}{
[( P_1^{n+2})^2 + H_1^{n+2}]^2 [( P_1^{n+3})^2 + H_1^{n+3}]} \\ \times|\psi_n(k_1,\dots,k_n)|
|\psi_n(k_2,\dots,k_n,k_{n+2})| dk_1\dots dk_{n+3} \\
\leq c_I^2 (\psi_n,|A|^*|A|\psi_n)\leq c_I^2 c_A (\psi_n,\Hf \psi_n),
\end{multline}
where we used $|\P| \leq \Hf$.

Finally we look at the
term $III_n$ where all indices differ, i.e.,
\begin{multline}
n^3 \int \frac{|\varphi(k_1)||\varphi(k_2)||\varphi(k_3)||\varphi(k_{n+1})||\varphi(k_{n+2})|
 |\varphi(k_{n+3})|| P_1^{n+2}||P_2^{n+3}|\times}{
[( P_1^{n+2})^2 + H_1^{n+2}]  [( P_2^{n+3})^2 + H_2^{n+3}] [( P_1^{n+3})^2 + H_1^{n+3}]} \\ \times|\psi_n(k_1,\dots,k_n)|
|\psi_n(k_4,\dots,k_{n+3})| dk_1\dots dk_{n+3} \\
\leq n^3 \int \frac{|\varphi(k_1)||\varphi(k_2)||\varphi(k_3)||\varphi(k_{n+1})||\varphi(k_{n+2})| |\varphi(k_{n+3})|\times}
{
[H_1^{n+2}]^{1/2}  [H_2^{n+3}]^{1/2}  H_1^{n+3}} \\ \times|\psi_n(k_1,\dots,k_n)|
|\psi_n(k_4,\dots,k_{n+3})| dk_1\dots dk_{n+3}\\ \leq
(|\psi_n|,|A|^* \Hf^{-1/2} |A|^* \Hf^{-1/2} |A|^*|A| \Hf^{-1/2}|A| \Hf^{-1/2}|\psi_n|) \leq c_A^3 (\psi_n,\Hf\psi_n).
\end{multline}
\end{proof}

\begin{lem}\label{she2b}
\begin{multline}
(\Psi,AA \Di \P A^* \Di A \P\Di A^*A^*\Psi) \\ \leq (\o,AA \Di \P A^* \Di A \P\Di A^*A^*\o) \|\Psi\|^2 + c (\Psi,\D \Psi)
\end{multline}
with $c$ uniformly bounded in $\Lambda$.
\end{lem}
\begin{proof}
The diagonal term looks like
\begin{multline}
I_n=2\int \frac{[\varphi(k_1)\c \varphi(k_2)][\varphi(\bar k_1)\c \varphi(k_2)][P_1^{n+1}\c \varphi(k_1)]
[\bP_1^{n+1}\c \varphi(\bar k_1)]\times}
{[(P_1^{n+2})^2 + H_1^{n+2}][(\bP_1^{n+2})^2 + \bH_1^{n+2}][(P_2^{n+2})^2 + H_2^{n+2}]}\\
\times |\psi_n(k_3,\dots,k_{n+2})|^2 dk_1 d\bar k_1 dk_2\dots dk_{n+2},
\end{multline}
where $\bP_1^l = \bar k_1+\sum_{i=2}^l k_i$ and $\bH_1^l = |\bar
k_1| + \sum_{i=2}^l |k_i|$.

We decompose as in \eqref{A16} and the main part
is estimated like
\begin{multline}
2 \int  \frac{[\varphi(k_1)\c \varphi(k_2)][\varphi(\bar k_1)\c
\varphi(k_2)][(P_1^{2}+ P_3^{n+2})\c \varphi(k_1)][(\bP_1^{2}+
P_3^{n+2})\c \varphi(\bar k_1)]\times} {[(P_1^{2})^2 +
H_1^{2}][(\bP_1^{n+2})^2 + \bH_1^{n+2}][(P_2^{2})^2 + H_2^{2}]}\\
\times |\psi_n(k_3,\dots,k_{n+2})|^2 dk_1 d\bar k_1 dk_2\dots
dk_{n+2} \\ \leq (\o,AA \Di \P A^* \Di A \P\Di A^*A^*\o)
\|\psi_n\|^2 \\+ c_{II}^3(\psi_n,\P^2\psi_n) + c_I
c_{II}^2(\psi_n,\Hf\psi_n).
\end{multline}
The remaining terms of the diagonal part are bounded analogously
to the error terms in the previous inequality, whereas the
off-diagonal terms are estimated like in the previous lemmas.
\end{proof}

\begin{lem}\label{she3}
\begin{multline}
(\Psi,AA \Di A^*A \Di A^*A^*\Psi) \\ \geq (\o,AA \Di A^*A \Di A^*A^*\o) \|\Psi\|^2 
- c (\Psi,\D \Psi) 
\end{multline}
with $c$ uniformly bounded in $\Lambda$.
\end{lem}
\begin{proof}
Since we now look for a lower bound, we
have to be a little bit more careful when treating the diagonal part
\begin{multline}
I_n= 2 \int \frac{[\varphi(k_1)\c \varphi(k_2)][\varphi(\bar
k_1)\c \varphi(k_2)][\varphi(\bar k_1)\c \varphi(k_1)]}
{[(P_1^{n+2})^2 + H_1^{n+2}][(\bP_1^{n+2})^2 + \bH_1^{n+2}]
}\times \\ \times |\psi_n(k_3,\dots,k_{n+2})|^2 d\bar k_1 dk_1
\dots dk_{n+2}\\ = 2 \int \frac{[\varphi(k_1)\c
\varphi(k_2)][\varphi(\bar k_1)\c \varphi(k_2)] [\varphi(\bar
k_1)\c \varphi(k_1)]}{[Q + b][\bar Q + \bar b] }\times \\ \times
|\psi_n(k_3,\dots,k_{n+2})|^2 d\bar k_1 dk_1 \dots dk_{n+2},
\end{multline}
with
$$Q = (P_1^{2})^2 + H_1^{2}, \quad b = (P_3^{n+2})^2 + H_3^{n+2} + 2 P_1^{2}\c P_3^{n+2}$$
and the equivalent expression for $\bar Q, \bar b$ replacing $k_1 $ by $\bar k_1$.
Using
\begin{multline}
\frac 1{Q +b} \frac 1{\bar Q +\bar b} = \left[ \frac 1{Q} - \frac{b}{Q(Q+b)}\right]\left[ \frac 1{\bar Q} -
\frac{\bar b}{\bar Q(\bar Q+\bar b)}\right]
\\ =  \frac 1{Q\bar Q} - \frac b{Q\bar Q(Q+b)}- \frac{\bar b}{\bar Q Q(\bar Q+\bar b)} + \frac{b \bar b}
{\bar Q Q(\bar Q+\bar b)(Q+b)},
\end{multline}
due to the symmetry of the two terms in the middle, we get immediately
\begin{multline}
I_n \geq
 2 \int \frac{[\varphi(k_1)\c \varphi(k_2)][\varphi(\bar k_1)\c \varphi(k_2)][\varphi(\bar k_1)\c 
 \varphi(k_1)]}{Q\bar Q }
 \times \\ \times
|\psi_n(k_3,\dots,k_{n+2})|^2 d\bar k_1 dk_1 \dots dk_{n+2} \\ - 6
 \int \frac{ |\varphi(k_1)|^2| \varphi(k_2)|^2|\varphi(\bar k_1)|^2 |b| }{Q\bar Q(Q+b) }
|\psi_n(k_3,\dots,k_{n+2})|^2 d\bar k_1 dk_1 \dots dk_{n+2}\\ \geq
(\o,AA \Di A^*A \Di A^*A^*\o) \|\psi_n\|^2 \\-  c_{II}^3
(\psi_n,\D \psi_n) -  c_{I}^2 c_{II}(\psi_n,\Hf\psi_n).
\end{multline}
Concerning $II_n$ we obtain two types of terms, namely
\begin{multline}
n \int \int \frac{|\varphi(k_{n+2})|^2||\varphi(k_1)||\psi_n(k_2,\dots,k_{n+1})|}{H_1^ {n+2}}dk_{n+2}\times \\
\int \frac{|\varphi(k_{1})||\varphi(k_2)| |\varphi(\bar k_{n+2})||\psi_n(k_3,\dots,k_{n+1},\bar k_{n+2})|}
{\bar H_1^ {n+2}}dk_{n+2}
dk_1 dk_{n+1}\\ \leq c_{II}^2 (|\psi_n|,|A|^*|A||\psi_n|) \leq c_{II}^2 c_A (\psi_n,\Hf \psi_n)
\end{multline}
and
\begin{multline}
n \int  \frac{|\varphi(k_1)|^2 |\varphi(k_2)|^2 |\varphi(k_{n+2})|
|\varphi(\bar k_{n+2})|
 |\psi_n(k_3,\dots,k_{n+1},\bar k_{n+2})|}{H_1^ {n+2}\bar H_1^ {n+2}} \times
\\ \times
|\psi_n(k_3,\dots, k_{n+2})||\psi_n(k_3,\dots,k_{n+1},\bar
k_{n+2}) | dk_1\dots dk_{n+2}d \bar k_{n+2}\\ \leq c_{II}^2
(|\psi_n|,|A|^*|A||\psi_n|) \leq c_{II}^2 c_A (\psi_n,\Hf \psi_n).
\end{multline}
Concerning $III_n$ we estimate
\begin{multline}
n^2 \int  \frac{|\varphi(k_1)|| \varphi(k_2)||\varphi(k_{n+2})||\psi_n(k_3,\dots,k_{n+2})|}{H_1^ {n+2}}\times \\
\times \frac{|\varphi(k_{1})||\varphi(k_{n+1})| |\varphi(\bar k_{n+2})||\psi_n(k_2,\dots,k_{n+1},\bar k_{n+2})|}
{\bar H_1^ {n+2}}
dk_1 \dots dk_{n+2} d\bar k_{n+2}\\ \leq c_{II} (|\psi_n|,|A|^*\Hf^{-1/2}|A|^*|A|\Hf^{-1/2}|A||\psi_n|) \leq c_{II} c_A^2
 (\psi_n,\Hf \psi_n)
\end{multline}
as well as concerning $IIII_n$
\begin{multline}
n^3 \int  \frac{|\varphi(k_1)|| \varphi(k_2)||\varphi(k_{n+2})||\psi_n(k_3,\dots,k_{n+2})|}{H_1^ {n+2}}\times \\
\times \frac{|\varphi(k_{n})||\varphi(k_{n+1})| |\varphi(\bar k_{n+2})||\psi_n(k_1,\dots,k_{n-1},\bar k_{n+2})|}
{\bar H_1^ {n+2}}
dk_1 \dots dk_{n+2} d\bar k_{n+2}\\ \leq
 (|\psi_n|,|A|^*\Hf^{-1/2}|A|^*\Hf^{-1/2}|A|^*|A|\Hf^{-1/2}|A|\Hf^{-1/2}|A||\psi_n|)\\
 \leq  c_A^3  (\psi_n,\Hf \psi_n).
\end{multline}
\end{proof}
The next Lemma is similar to the ones explained in \cite[Appendix C]{CH}.
\begin{lem}\label{she4}
\begin{eqnarray}
&(i)& |(\Psi,AA \Di \P A \Di\P A \Di A^*A^* \Psi)| \leq c\|\Psi\|\|\Hf^{1/2}\Psi\|,\\
&(ii)& |(\Psi,AA \Di \P A \Di A^*A^* \Psi)| \leq c \|\Psi\|\|\Hf^{1/2}\Psi\|
\end{eqnarray}
with $c$ uniformly bounded in $\Lambda$.
\end{lem}
\begin{proof}
We sketch the proof of (ii). (i) works analogously.
The diagonal part reads
\begin{multline}
n\int \frac {\varphi(k_n) \c \varphi(k_{n+1})P_1^{n+1}\c \varphi(k_{n+2})\overline{\psi_{n-1}(k_1,\dots,k_{n-1})}}
{(P_1^{n+1})^2 + H_1^{n+1}}\times \\
\times \frac{\varphi(k_{n+1}) \c \varphi(k_{n+2})\psi_{n}(k_1,\dots,k_{n})}{(P_1^{n+2})^2 + H_1^{n+2}} \\
\leq c_Ic_{II} (|\psi_{n-1}|,|A||\psi_n|) \leq c_Ic_{II}c_A^{1/2} \|\psi_{n-1}\|\|\Hf^{1/2}\psi_n\|.
\end{multline}
By methods similar to the previous lemmas
the off-diagonal terms are estimated by
$(|A||\psi_{n-1}|,|A|\Hf^{-1/2}|A||\psi_n|)$, respectively by\\
$(|A|\Hf^{-1/2}|A||\psi_{n-1}|,|A|\Hf^{-1/2}|A|\Hf^{-1/2}|A||\psi_n|)$.
\end{proof}

\section{Sharp Estimates needed for Theorem \ref{thm2}}
\label{Bap}

We introduce
\begin{equation}
c(e) = \int \frac{|\varphi(k)|^2}{|k|[|k|+e^7]} dk \leq c_{II} \ln[1/e].
\end{equation}
Furthermore recall that for all $0 \leq \eps < 1$
\begin{equation}\label{epsi}
\eps P^2 \leq (P^2 + V  -\eat) + \eps |\eat|/(1-\eps),
\end{equation}
from which we obtain
\begin{equation}\label{P2B}
 P^2 \leq 2 (P^2 + V - \eat) +  c
\end{equation}
with $c=2|\eat|$. Inserting $\eps = \Hf/(\Hf - \eat)$ in \eqref{epsi} 
(cf. \cite[Equation (4.16)]{HS}) shows
\begin{equation}\label{PBP}
 P \frac 1{P^2 + V - \eat +  \Hf}P \leq 1 + \frac {|\eat|}\Hf
\end{equation}
as well as 
\begin{equation}\label{BP2B}
\frac{1}{P^2 + V - \eat + \Hf} P^2 \frac{1}{P^2 + V - \eat + \Hf} \leq c\left(\frac 1\Hf + \frac 1{\Hf^2}\right).
\end{equation}
In the following we deal with states  of the form $h\vp\c \vp$, meaning
we understand that as
\begin{equation}\label{B6}
[h\vp\c \vp]_{n+2}(x,k_1,\dots,k_{n}) = h_n(x,k_1,\dots,k_{n})\vp(k_{n+1})\c \vp(k_{n+2}),
\end{equation}
where $h \in \H$. This wave function
is introduced for notational simplification. It is not symmetric in all variables.
This does not matter, since all operations also hold for 
the case of general wave functions, once we 
extend the definition of $A$ as
\begin{equation}
[A \Psi]_{n-1}(x,k_1,\dots,k_{n-1})=\frac 1{\sqrt{n}} \sum_{i=1}^n \int \varphi(k_i) \psi_n(x,k_1,\dots,k_i,\dots,k_n)dk_i.
\end{equation}
Recall that
\begin{equation*}
B= P^2 + V -\eat + \Hf.
\end{equation*}

\begin{lem}\label{alem2}
\begin{equation}
(h,AAB^{-1} A^*A^*h) \leq  (h\vp\c\vp,B^{-1} h\vp\c\vp) + c(h,\Hf h),
\end{equation}
with 
\begin{equation}\label{alem2cc}
(h\vp\c\vp,B^{-1} h\vp\c\vp)\leq c\|h\|^2,
\end{equation}
where the constants are uniformly bounded in 
the cutoff.
\end{lem}
\begin{proof}
We fix again a photon number $n$.
Recall, as noted in \eqref{defDD},
\begin{multline}\label{defDDe}
[A^* A^*h_n]_{n+2} = \frac 1{\sqrt{(n+2)(n+1)}}
 \sum_{j=1}^{n+2} \sum_{\substack{i=1 \\ i \neq j}}^{n+2}
\varphi(k_j)\cdot \varphi(k_i) \times\\\times
h_n(k_1,\dots,\not\!\! k_j,\dots,\not \!\! k_i,\dots,k_{n+2}).
\end{multline}
By symmetry we again distinguish
three different terms, where the first,
diagonal term, is simply given
as 
\begin{equation}
(h_n\vp\c\vp,B^{-1} h_n\vp\c\vp).
\end{equation}
This is the term desired and
we only need to estimate the 
off-diagonal terms.
We proceed in analogy to 
$II_n$ in Lemma \ref{she1}.
Namely, 
\begin{multline}\label{tr1}
n\big(h_n(x,k_1,\dots,k_n)\vp(k_{n+1})\c\vp(k_{n+2}),B^{-1}\vp(k_{1})\c \vp(k_{n+2}) h_n(x,k_2,\dots,k_{n+1})\big)\\
= n \Big(h_n(x,k_1,\dots,k_n)|k_1|^{1/2}\frac{\vp(k_{n+1})}{|k_{n+1}|^{1/2}}\c\vp(k_{n+2}),B^{-1}\times \\ \times
\frac{\vp(k_{1})}{|k_1|^{1/2}}\c \vp(k_{n+2})|k_{n+1}|^{1/2} h_n(x,k_2,\dots,k_{n+1}\Big) \\ \leq
c_{II}^2 (h_n,\Hf h_n).
\end{multline}

The second off-diagonal term is given by
\begin{multline}
n^2 \Big(\varphi(k_1) \c \varphi(k_2)h_n(x,k_3,\dots,k_{n+2}),{B}^{-1} h_n(x,k_1,\dots,k_{n})\varphi(k_{n+1})\c
\varphi(k_{n+2})\Big).
\end{multline}
We rewrite it as
\begin{multline}\label{n2schw}
n^2\Big(\frac{ \varphi(k_1)\c
\varphi(k_2)}{|k_1|^{1/2}|k_2|^{1/2}}|k_{n+1}|^{1/2}|k_{n+2}|^{1/2}h_n(x,k_3,\dots,k_{n+2}),
\frac 1{B} \times \\ \times|k_1|^{1/2}|k_2|^{1/2} h_n(x,k_1,\dots,k_{n})
\frac{\varphi(k_{n+1})\c \varphi(k_{n+2})}{|k_{n+1}|^{1/2}|k_{n+2}|^{1/2}}\Big) \\
\leq
n^2\int \frac{ |\varphi(k_1)|^2
|\varphi(k_2)|^2}{|k_1||k_2|}dk_1 dk_2 \int
|k_{n+1}||k_{n+2}|\frac{|h_n(x,k_3,\dots,k_{n+2})|^2}{\sum_{i=3}^{n+1}
|k_i|}dk_3...dk_{n+2}\\ \leq C n \int
|k_{n+2}|\sum_{i=3}^{n+1}
|k_i|\frac{|h_n(x,k_3,\dots,k_{n+2})|^2}{\sum_{i=3}^{n+1}
|k_i|}dk_3...dk_{n+2} \\ \leq C n \int |k_{n+2} ||h_n|^2 dk_3...dk_{n+2}= C (h_n, \Hf h_n),
\end{multline}
where we used Schwarz inequality, the fact that $1/B \leq 1/\Hf$,
and the symmetry of $h_n(x,k_3,...,k_{n+2})/(\sum_{i=3}^{n+1}
|k_i|)$ in the variables $k_3$ to $k_{n+1}$.
Obviously \eqref{alem2cc}
holds since $(h\vp\c\vp,B^{-1} h\vp\c\vp) \leq c_I^2$.
\end{proof}

\begin{lem}\label{alem3}
\begin{multline}
\big | (h, P A B^{-1} P A B^{-1} A^* A^* h) \\
- (\vp  \c P B^{-1} \vp\c Ph,  B^{-1}  h \vp\c \vp) \big| \leq  c(e)^{1/2} \|Ph\|\|\Hf^{1/2}h\|
\end{multline}
with 
\begin{equation}\label{alem2cd}
|(\vp  \c P B^{-1} \vp\c P h,  B^{-1}  h \vp\c \vp )| \leq c\|h\|\|Ph\|.
\end{equation}
\end{lem}
\begin{proof}
We can estimate the first off-diagonal term by using Schwarz inequality and by a similar  
calculation as in \eqref{tr1},
\begin{multline}\label{b16}
n( P  \c \vp(k_{n+2}) B^{-1} P\c  \vp(k_{n+1})  h_n(x,k_1,\dots,k_n) , B^{-1}\times \\ \times \vp(k_{1})\c \vp(k_{n+2}) h_n(x,k_2,\dots,k_{n+1}))
\\ \leq\Big[ n\Big (\vp(k_{n+2})\c P B^{-1} \frac{\vp(k_{n+1})}{|k_{n+1}|^{1/2}}\c P h_n(x,k_1,\dots,k_n)|k_1|^{1/2}, |k_{n+2}|^{-1} \times \\ \times 
\vp(k_{n+2})\c P B^{-1}  \frac{\vp(k_{n+1})}{|k_{n+1}|^{1/2}}
\c P h_n(x,k_1,\dots,k_n)|k_1|^{1/2}\Big)\Big]^{1/2}
 \big[c_{II}^2 (h_n,\Hf h_n)\big]^{1/2}
\\ \leq \Big[  c_{II}n\Big(P h_n(x,k_1,\dots,k_n)|k_1|^{1/2} \frac{|\vp(k_{n+1})|}{|k_{n+1}|^{1/2} } ,
[\Hf^{-1} +\Hf^{-2}] \times \\ \times \frac{|\vp(k_{n+1})|}{|k_{n+1}|^{1/2} }P h_n(x,k_1,\dots,k_n)|k_1|^{1/2}\Big)\Big]^{1/2}\big[c_{II}^2 (h_n,\Hf h_n)\big]^{1/2} \\
\leq \left[  c_{II} c(e)n\Big(Ph_n,\frac{|k_1|}{\sum_{i=1}^n|k_i|}Ph_n\Big)\right]^{1/2}\big[c_{II}^2 (h_n,\Hf h_n)\big]^{1/2}\\
\leq  c(e)^{1/2} \|Ph_n\|\|\Hf^{1/2}h_n\|,
\end{multline}
where also \eqref{BP2B} is used.
For the second off-diagonal term we proceed similarly.
Thereby, after Schwarz inequality, the more
difficult term, suppressing the square root, can be bounded by
\begin{multline}\label{b17}
n^2\Big (Ph_n (x,k_1,\dots,k_n) |k_1|^{1/2}|k_2|^{1/2} \frac{|\vp(k_{n+1})|}{|k_{n+1}|^{1/2} } \frac{|\vp(k_{n+2})|}{|k_{n+2}|^{1/2} } \big[\Hf^{-1} + \Hf^{-2}]
\times \\ \times \Hf^{-1} Ph_n (x,k_1,\dots,k_n) |k_1|^{1/2}|k_2|^{1/2} \frac{|\vp(k_{n+1})|}{|k_{n+1}|^{1/2} } \frac{|\vp(k_{n+2})|}{|k_{n+2}|^{1/2} }\Big)\\
\leq c n^2 (Ph_n,\frac{ |k_1||k_2|}{(\sum_{i=1}^n|k_i|)^2}Ph_n)  \leq c \|Ph_n\|^2,
\end{multline}
where we used $B\geq \Hf$ and \eqref{BP2B}.
The inequality \eqref{alem2cd} is obvious.
\end{proof}
\begin{lem}\label{alem3b}
\begin{multline}
\big | (h, A  A B^{-1}A^*A  B^{-1} A^* A^* h)  \\- (h \vp \c \vp,B^{-1}A^*A  B^{-1}  h \vp \c \vp) \big| \leq c(h,\Hf h)
\end{multline}
with 
\begin{equation}
(h\vp \c \vp,B^{-1}A^*A  B^{-1}  h \vp \c \vp)  \leq c\|h\|^2.
\end{equation}
\end{lem}
\begin{proof}
Denote $S=B^{-1/2}A^*A  B^{-1/2}$.
Notice, due to Lemma \ref{hlt2}, $\| A  B^{-1/2}\| \leq c_A^{1/2}$ and consequently
$S\leq c$. 
The result follows by applying the proof of Lemma \ref{alem2} to $(h, A  A B^{-1/2}S B^{-1/2} A^* A^* h) .$
\end{proof}

\begin{lem}\label{alem4}
\begin{multline}
(h, A A B^{-1} P A B^{-1}A^*  P B^{-1} A^* A^* h)\\ \leq
(h\vp \c \vp,B^{-1} P A B^{-1}A^*  P B^{-1}h\vp \c \vp) + (c+c(e)) (h,\Hf h),
\end{multline}
with 
\begin{equation}\label{alem4d}
(h\vp \c \vp,B^{-1} P A B^{-1}A^*  P B^{-1}h\vp \c \vp) \leq c\|h\|^2.
\end{equation}
\end{lem}
\begin{proof}
Denote $S=B^{-1} P A B^{-1}A^*  P B^{-1}$.
Using Lemma \ref{hlt2}  together with \eqref{BP2B}
we see that $S\leq c[\Hf^{-1} + \Hf^{-2}]$.
The off-diagonal terms can be estimated by using $S$.
The terms corresponding to $\Hf^{-1}$ are treated as in Lemma \ref{alem2},
whereas the terms corresponding to $\Hf^{-2}$ 
can be bounded by similar methods (cf., e.g., the calculations of \eqref{b16} and \eqref{b17}) by $c(e)(h_n,\Hf h_n)$
with fixed but arbitrary photon number.
\end{proof}
\begin{lem}\label{alem4b}
\begin{multline}
(h, A A B^{-1}  A^* P B^{-1}   P A B^{-1} A^* A^* h)\\ \leq
(h\vp \c \vp,B^{-1} A^* P  B^{-1} P A B^{-1}h\vp \c \vp) + (c+c(e)) (h,\Hf h)
\end{multline}
with 
\begin{equation}
(h\vp \c \vp,B^{-1} A^*PB^{-1}  P A  B^{-1}h\vp \c \vp) \leq c\|h\|^2.
\end{equation}
\end{lem}
\begin{proof}
The proof proceeds analogously to Lemma \ref{alem4},
since according to Lemma \ref{hlt2} $$B^{-1} A^*PB^{-1}  P A  B^{-1} \leq c [\Hf^{-1} + \Hf^{-2}]$$ also holds.
\end{proof}

\begin{lem}\label{alem5}
\begin{equation}
|(\Psi,AAB^{-1}PAB^{-1}A^*A^*\Psi)| \leq c(e)\|\Psi\|\|\Hf^{1/2}\Psi\|.
\end{equation}
\end{lem}
\begin{proof}
The proof proceeds similarly to Lemma \ref{she4} and \cite[Appendix C]{CH}.
We demonstrate it on the ``diagonal'' term
\begin{multline}\label{B24}
\sqrt{n} \big( \vp(k_{n+2}) \c P B^{-1}\vp(k_n)\c \vp(k_{n+1})\psi_{n-1}(x,k_1,\dots,k_{n-1}), B^{-1} \times \\ \times
\psi_n(x,k_1,\dots,k_{n}) \vp(k_{n+1})\c \vp(k_{n+2} )\big) \\ =
\sqrt{n}\Big  (\vp(k_{n+2}) \c P B^{-1}\frac{\vp(k_n)}{|k_n|^{1/2}} \c \vp(k_{n+1})\psi_{n-1}(x,k_1,\dots,k_{n-1}), B^{-1} \times \\
\times
\psi_n(x,k_1,\dots,k_{n})|k_{n}|^{1/2} \vp(k_{n+1})\c \vp(k_{n+2} )\Big).
\end{multline}
Using \eqref{BP2B}, $B \geq |k_{n+2}|$ together with Schwarz inequality,
we bound 
\begin{multline}
|\eqref{B24}| \leq \Big[c_{II} \Big(\psi_{n-1}(x,k_1,\dots,k_{n-1})\frac{|\vp(k_n)|}{|k_n|^{1/2}}| \vp(k_{n+1})|,[\Hf^{-1} + \Hf^{-2}] \times  \\
\times \psi_{n-1}(x,k_1,\dots,k_{n-1})\frac{|\vp(k_n)|}{|k_n|^{1/2}}| \vp(k_{n+1})|\Big)\Big]^{1/2} c_I (\psi_n,\Hf\psi_n)^{1/2}
\\ \leq c_I c_{II} c(e)^{1/2} \|\psi_{n-1}\| \|\Hf^{1/2} \psi_n\|.
\end{multline}
The remaining terms are covered in a similar fashion.
\end{proof}

\section{Operator inequalities}
\label{Cap}

In this section we state and prove some operator
inequalities used in the proof of Theorem \ref{thm1} and \ref{thm2}.

We start with a simple but useful Lemma for our
estimates employed in the proof of Theorem \ref{thm1}.
\begin{lem}
\label{hlt1} In the sense of forms
we have
\begin{eqnarray}
(i) && A\Di A^* \leq c,\\
(ii) && AA \Di A^*A^* \leq c(1+\Hf).
\end{eqnarray}
Since $L \geq \D$ the above inequalities
also hold for $L$.
\end{lem}
\begin{proof}
$(i)$
\begin{equation}
\|\Dh A^*\| = \|A\Dh\| \leq \|A\Hf^{-1/2}\|\leq c_A^{1/2}.
\end{equation}
$(ii)$ Follows directly from
the proof of Lemma \ref{she1},
since $I_n$ in \eqref{A7}
can be bounded by
$c_I^2 \|\psi_n\|^2$.
\end{proof}
The auxiliary Lemma for the proof
of Theorem \ref{thm2} is a bit more involved.

\begin{lem}
\label{hlt2} In the sense of forms we have
\begin{eqnarray*}
(i) && A B^{-1} A^* \leq c,\\
(ii) && A B^{-2} A^* \leq c(e),\\
(iii) && AA B^{-1} A^*A^* \leq c(1+\Hf),\\
(iv) &&  AA B^{-2} A^*A^* \leq c,\\
(v) && AA B^{-1}P^2 B^{-1}  A^*A^* \leq c(1+\Hf),\\
(vi) && A B^{-1}P^2 B^{-1}  A^* \leq c(e)+c\\
(vii) && A B^{-1} A^* \Hf^{-1} A  B^{-1} A^* \leq c \Hf^{-1}
\end{eqnarray*}
Since $K \geq B$, the above inequalities
also hold for $K$.
\end{lem}
\begin{proof}
$(i)$ is a simple consequence of Lemma \ref{hlt1} $ (i)$,
since $B \geq \Hf$.

\noindent
$(ii)$ Observe $A B^{-2} A^* \leq A \Hf^{-2} A^*$.
The corresponding diagonal part is bounded by
\begin{multline}
\big(|\varphi(k_1)||\psi_n(x,k_2,\dots,k_{n+1})|,\Hf^{-2}|\varphi(k_1)||\psi_n(x,k_2,\dots,k_{n+1})|\big)
\\ \leq \|\psi_n\|^2 \int \frac{|\vp(k)|^2}{[|k|+e^7]^2} dk \leq c(e) \|\psi_n\|^2.
\end{multline}
The off-diagonal part is estimated by
\begin{multline}
n \big(|\varphi(k_1)||\psi_n(x,k_2,\dots,k_{n+1})|,\Hf^{-2}|\psi_n(x,k_1,\dots,k_n)||\varphi(k_{n+1})|\big)
\\ \leq n \Big(\frac{|\varphi(k_1)|}{|k_{n+1}|^{1/2}}|k_{n+1}|^{1/2}|\psi_n(x,k_2,\dots,k_{n+1})|,\times \\ \times
\Hf^{-2}|\psi_n(x,k_1,\dots,k_n)||k_{1}|^{1/2}\frac{|\varphi(k_{n+1})|}{|k_{n+1}|^{1/2}}\Big)
\\ \leq
 \int \frac{|\vp(k)|^2}{|k|[|k|+e^7]} dk \,n(\psi_n,\frac{|k_1|}{\sum_{i=1}^n|k_i| }\psi_n) \leq c(e) \|\psi_n\|^2.
\end{multline}
$(iii)$ is obvious.

\noindent
$(iv)$ The first two terms
are treated similarly to $(ii)$,
only this time we have the finite bounds $c_{II}^2$ thanks to the fact there are two photons created.

The third term, where the indices in the created photons as well
as in the wave function $\psi_n$ are distinct,
is estimated by
\begin{multline}
 n^2\Big(\frac{ |\varphi(k_1)||
\varphi(k_2)|}{|k_1|^{1/2}|k_2|^{1/2}}|k_{n+1}|^{1/2}|k_{n+2}|^{1/2}|\psi_n(x,k_3,\dots,k_{n+2})|,
\frac 1{\Hf^2} \times \\ \times|k_1|^{1/2}|k_2|^{1/2}| \psi_n(x,k_1,\dots,k_{n})|
\frac{|\varphi(k_{n+1})||\varphi(k_{n+2})|}{|k_{n+1}|^{1/2}|k_{n+2}|^{1/2}}\Big) \\ \leq c_{II}^2 n \int
|k_{n+2}|\sum_{i=3}^{n+1}
|k_i|\frac{|\psi_n(x,k_3,\dots,k_{n+2})|^2}{(\sum_{i=3}^{n+2}
|k_i|)^2}dk_3...dk_{n+2} \\ \leq c_{II}^2 n \int \frac{|k_{n+2} |}{\sum_{i=3}^{n+2}
|k_i|}|\psi_n|^2 dk_3\dots dk_{n+2} = c_{II}^2 \|\psi_n\|^2.
\end{multline}
Observe that by means of \eqref{BP2B} together with Lemma \ref{hlt2} $(i) -(iv)$
we arrive at $(iv)$ and $(v)$.

\noindent $(vii)$ is an easy application
of our method and can be guessed immediately,
since $ A  B^{-1} A^* \leq c$.

\end{proof}
By means of  Lemma \ref{hlt1}
we can easily prove  the operator inequalities 
used in the proof of  Theorem \ref{thm2}.

\begin{lem}
\label{hlt3}
\begin{eqnarray*}
(i) && \|B^{-1/2} A^*A^* B^{-1} A^*\Psi\| \leq \|\Psi\|(c + c(e))^{1/2} \\
(ii) && \|B^{-1/2} A^*A^* B^{-1} A^*A^*\Psi\| \leq \|(1 + \Hf)^{1/2}\Psi\|,\\
(iii) &&\|B^{-1/2} A^*P B^{-1} A^*\Psi\| \leq \|\Psi\|(c + c(e))^{1/2} ,\\
(iv) && \|B^{-1/2} A^*P B^{-1} A^*A^*\Psi\| \leq \|(1 + \Hf)^{1/2}\Psi\| ,\\
(v) && \|B^{-1/2} PA B^{-1} A^*A^*\Psi\| \leq \|(1 + \Hf)^{1/2}\Psi\|\\
(vi) && \|B^{-1/2} A^*P B^{-1} A^*A^*B^{-1}A^*\Psi\| \leq \|\Psi\|(c + c(e))^{1/2}\\
(vii) && \|B^{-1/2} A^*P B^{-1} A^*A^*B^{-1}A^*A^*\Psi\| \leq \|(1 + \Hf)^{1/2}\Psi\| .\\
\end{eqnarray*}
Since $K \geq B$, the above inequalities
also hold for $K$.
\end{lem}
\begin{proof}
$(i)$ and $(ii)$ are a simple consequence
of Lemma \ref{hlt2} $(i)-(iv)$.

\noindent
For $(iii)$ and $(iv)$ apply Lemma \ref{hlt2} $(i)$ and
$(v)$, respectively $(vi)$.

\noindent
For $(v)$ apply
$P B^{-1}P \leq c(1 + \Hf^{-1})$.
Furthermore use Lemma \ref{hlt2} $(vii)$.
This together with Lemma \ref{hlt2} $(iii)$ and $(iv)$ implies the inequality.

\noindent
$(vi)$ and $(vii)$ are a direct consequence
of Lemma \ref{hlt2}.
\end{proof}

\end{appendix}

\end{document}